# The empirical likelihood prior applied to bias reduction of general estimating equations

Albert Vexler, Li Zou

*Department of Biostatistics, State University of New York at Buffalo, Buffalo,*

*New York 14214, U.S.A.*

avexler@buffalo.edu            lizou@buffalo.edu

and

Alan D. Hutson

*Department of Biostatistics and Bioinformatics, Roswell Park Cancer Institute, Buffalo, New*

*York 14263, U.S.A.*

alan.hutson@roswellpark.org

**ABSTRACT**

The practice of employing empirical likelihood (EL) components in place of parametric likelihood functions in the construction of Bayesian-type procedures has been well-addressed in the modern statistical literature. We rigorously derive the EL prior, a Jeffreys-type prior, which asymptotically maximizes the Shannon mutual information between data and the parameters of interest. The focus of our approach is on an integrated Kullback-Leibler distance between the EL-based posterior and prior density functions. The EL prior density is the density function for which the corresponding posterior form is asymptotically negligibly different from the EL. We show that the proposed result can be used to develop a methodology for reducing the asymptotic bias of solutions of general estimating equations and M-estimation schemes by removing the first-order term. This technique is developed in a similar manner to methods employed to reduce the asymptotic bias of maximum likelihood estimates via penalizing the underlying parametric likelihoods by their Jeffreys invariant priors. A real data example related to a study of myocardial infarction illustrates the attractiveness of the proposed technique in practical aspects.

# 1     Introduction

It is well-known that in order to carry-forth Bayesian inference a prior density and a likelihood function need to be specified. The selection and justification of priors are important components of Bayesian methodology. A widely used prior distribution was proposed by Jeffreys (1946). In this fundamental work, Jeffreys employed the Kullback-Leibler (K-L) measure to quantify a distance between the corresponding posterior and prior density functions. An excellent review of the justifications for Jeffreys prior is presented in Hartigan (1964, 1998). In parametric Bayesian inference, great efforts were made to formulate prior distributions, which add minimum information to knowledge derived from data (Bernardo 1979). Prior distribution functions are defined as reference priors if: 1) They roughly describe situations in which little relevant information is available; and 2) The corresponding posterior distributions provide a standard in which other distributions could be referred to in order to assess the relative importance of the initial knowledge in the final results (Bernardo 1979; Berger, Bernardo, and Sun 2009). Bernardo (1979) presented a heuristic discussion of the basic ideas related to the development of reference prior distributions that maximize Shannon mutual information between the posterior and prior density functions. In this case, under regularity conditions, the Jeffreys prior was shown to be a reference prior. Furthermore, these results were rigorously shown in asymptotic forms (see for details Lehmann and Casella 1998, pp. 261-262).

The above mentioned analysis related to the selection of priors corresponds to the parametric setting when the form of the likelihood is completely specified. Hartigan (1998) used a truncated K-L loss approach to develop maximum likelihood prior densities such that the corresponding Bayesian posterior functions are asymptotically negligibly different from the maximum likelihood functions. This approach offers a useful linkage between frequentist and Bayesian methods. In this framework Jeffreys prior is the unique continuous prior that yields the Bayesian strategy that asymptotically achieves the maximum Bayes risk in the context of a relevant entropy loss measure (Bernardo 1979; Lehmann and Casella 1998; Berger, Bernardo, and Sun 2009).

The Bayesian principle is one of the central tenets for developing powerful statistical inference tools when the form of the data distributions are assumed to be known and certain key assumptions are met. These principles may not be fiducial and applicable in the nonparametric setting when the likelihood function forms are assumed to be unknown. It is also well-known that when key parametric assumptions are not met, the parametric Bayesian approach may be suboptimal or biased (Daniels and Hogan 2008; Zhou and Reiter 2010). Towards this end, we can find within the modern applied and theoretical statistical literature a line of research around Bayesian empirical likelihood (BEL) techniques based on the empirical likelihood (EL) concept (Lazar 2003; Chaudhuri and Ghosh 2011; Yang and He 2012; Vexler, Ge, and Hutson 2014, 2016; Zhong and Ghosh 2016). Lazar (2003) theoretically justified that EL functions can be applied towards constructing nonparametric posterior distributions. In this context, EL's can provide valid posterior inference that satisfies the laws of probability in the sense that it is related to statements derived from the Bayes' rule (Lazar 2003). Vexler, Ge, and Hutson (2014) used EL functions to develop robust and efficient posterior point estimators. Furthermore, Vexler, Zou, and Hutson (2016) proposed and examined the BEL credible set estimation as an analogue to the traditional and efficient parametric Bayesian approach. Recently, Zhong and Ghosh (2016) provided an expression for the asymptotic expansion of posteriors that are based on EL component along with its variants. In general, the BEL method employs EL components in place of parametric likelihood functions in order to develop distribution-free Bayesian-type procedures.

The objectives in this paper are twofold: 1) To rigorously construct a Jeffreys-type EL prior in the context of the BEL algorithm; and 2) To develop a bias reduction approach for maximum empirical likelihood estimates (MELEs) by penalizing the EL by its prior. The second aim is inspired by the well-known fact that in the parametric setting Jeffreys priors can be used to penalize the underlying likelihood functions in order to reduce the bias of the corresponding maximum likelihood estimates (Firth 1992, 1993). MELEs can be associated with solutions of general estimating equations (Qin and Lawless 1994) as well as with estimates obtained by employing the generalized method of moments (Hansen 1982). In addition, the MELE framework can be extended to certain *M*-estimators (Owen 1988). It is well-known that bias corrected ELs inherit the higher order asymptotic efficiency properties of the maximum likelihood (Newey and Smith 2004). Thus, we provide the theoretical justification regarding the use of the derived EL prior in order to penalize the EL in an effort to improve the small sample properties of MELEs. This will be applied to reducing the EL bias in the framework of general estimating equations and *M*-estimation schemes.

This paper is organized as follows: In Section 2, we derive the EL prior and evaluate its asymptotic properties. In this section, we attend to necessary conditions related to the ability to consider BEL algorithms in the context of Shannon mutual information between data and parameters of interest. In Section 3, we demonstrate the process for using EL prior densities for the purpose of eliminating the first-order bias terms from the asymptotic expectation of MELEs. In Section 4, an extensive Monte Carlo (MC) study is conducted to examine the proposed method. The applicability of the proposed method is illustrated through a real world example of myocardial infarction in Section 5. In Section 6, we provide concluding remarks. Proofs of the theoretical results presented in this paper are outlined in Appendix A.

## 2 Empirical Likelihood Prior

To outline the central concept for constructing a reference prior in the classic parametric Bayesian setting and, without loss of generality, we assume that the data set *X* consists of *n* independent and identically distributed observations, $X=(X_1,\ldots,X_n)$, and $f(x|\theta)$ is the density function of $X_1$ with a scalar parameter $\theta$. In this case, the Bayesian posterior density has the form

$$\pi(\theta \mid X) = L(X \mid \theta)\pi(\theta) / \int L(X \mid \zeta)\pi(\zeta)d\zeta,$$

where $\pi(\theta)$ is the prior density function and $L(X \mid \theta)$ denotes the parametric likelihood $\prod_{i=1}^{n} f(X_i \mid \theta)$, provided that the form of $f(\cdot)$ is known. According to Lindley (1956) and Bernardo (1979), the prior density $\pi(\theta)$ is a reference prior if it maximizes the functional

$$I(\pi) = \int\int \pi(\theta) L(X \mid \theta) \log\{\pi(\theta \mid X)/\pi(\theta)\} dX d\theta \qquad (1)$$

in an asymptotic ($n \to \infty$) fashion. The information-theoretic quantity, $I(\pi)$, measures the amount of missing information about $\theta$ when the prior is fixed to be $\pi(\theta)$ (Bernardo, 1979). The use of the functional $I(\pi)$ allows one to make precise the basic idea of the construction of reference priors, which are maximally dominated by data. This concept was formalized in Berger, Bernardo, and Sun (2009). Bernardo (1979) showed that the arguments from the calculus of variations can be applied to approximate the reference prior in the form $\pi(\theta) \propto \exp\{\int L(X \mid \theta) \log\{\pi_A(\theta \mid X)\} dX\}$, where $\pi_A(\theta \mid X)$ denotes the corresponding asymptotic posterior density of $\theta$. Furthermore, it was proved that $I(\pi)$ satisfies

$$I(\pi) = \frac{1}{2}\log\frac{n}{2e\pi} + \int \pi(\theta) \log\left[\{i(\theta)\}^{0.5}/\pi(\theta)\right] d\theta + o(1), \text{ as } n \to \infty, \qquad (2)$$

where $i(\theta) = -E\{\partial^2 \log f(X_i \mid \theta)/\partial \theta^2\}$ is the Fisher information (see for details Lehmann and Casella 1998, pp. 261-262). Since the integral in the asymptotic form (2) of $I(\pi)$ is the only term involving the prior $\pi(\theta)$, maximizing that integral will maximize the asymptotic expansion. Jeffreys prior, $\pi(\theta) \propto \{i(\theta)\}^{0.5}$, is the appropriate choice that ensures the maximization of the integral. In accordance with Hartigan (1998), $I(\pi)$ can be represented in the form

$$I(\pi) = \int \pi(\theta) E_\theta[\log\{\pi(\theta \mid X)/\pi(\theta)\}] d\theta, \qquad (3)$$

where the operator $E_\theta[\cdot]$ denotes the mathematical expectation with respect to the underlying data density function $L(X \mid \theta)$. We refer the reader to Hartigan (1998, p. 2084) for more details regarding the definition (3). This statement was used in Hartigan (1998) to analyze the reference prior concept based on maximum likelihood functions. We extend this method to derive a higher order approximation related to the BEL approach in the context of Shannon's mutual information. Towards this end, we define the log EL function,

$$l(\theta) = \max_{0<p_1,\ldots,p_n<1} \left\{ \sum_{i=1}^{n} \log p_i : \sum_{i=1}^{n} p_i = 1, \sum_{i=1}^{n} G(X_i, \theta) p_i = 0 \right\}, \quad (4)$$

where $p_i$, $i=1,\ldots,n$, play the role of probability weights and the parameter of interest $\theta$ satisfies $E\{G(X_i, \theta)\} = 0$ for a specified function $G$. Then the EL function, $\exp\{l(\theta)\}$, will replace the parametric likelihood function $L(X \mid \theta)$ in the expression of the posterior density $\pi(\theta \mid X)$. In this section, we present necessary conditions and rigorous considerations for deriving the EL prior; see Remark A1 in the Supplementary Material (SM) for additional details (see Appendix B).

In general, EL functions cannot be directly used to construct a nonparametric counterpart of $I(\pi)$ defined in (3). To explain this issue, we consider a simple but commonly used EL for the mean. In this setting $G(X_i, \theta) = X_i - \theta$, $i=1,\ldots,n$. Let us convert the expression in equation (3) into its corresponding nonparametric form given as

$$I_e(\pi) = \int \pi(\theta) E[\log\{\pi_e(\theta \mid X) / \pi(\theta)\}] d\theta,$$

where $\pi_e(\theta \mid X) = \pi(\theta) \exp\{lr(\theta)\} / \int \pi(\zeta) \exp\{lr(\zeta)\} d\zeta$ and the log EL ratio $lr(\theta) = \log[\exp\{l(\theta)\} n^n]$. This form requires attention to the following technical issues: 1) The log EL ratio, $lr(\theta)$, is not defined for all values of $\theta$ that are considered in the integration process. This problem is associated with an ability to extract values of the probability weights $p_i$, $i=1,\ldots,n$, that maximize the EL function, $\prod_{i=1}^{n} p_i$, satisfying the constraints $\sum_{i=1}^{n} p_i = 1$ and $\sum_{i=1}^{n} X_i p_i = \theta$. It is clear that the probability weights ($p_i$'s) cannot be calculated when $\theta \notin \left\{\min_{i=1,\ldots,n}(X_i), \max_{i=1,\ldots,n}(X_i)\right\}$. The empirical bounds on $\theta$, i.e. $\min_{i=1,\ldots,n}(X_i)$ and $\max_{i=1,\ldots,n}(X_i)$, cannot be applied in the integration process, since the operator $E$ is in effect under the integral; 2) The formal notation $I_e(\pi)$ assumes that the expectation $E[\log\{\pi(\theta|X)/\pi(\theta)\}]$ exists, whereas one can show for certain values of $\theta$ this expectation might be unbounded (see Remark A2 in the SM, Appendix B).

Towards this end, we first consider the definition of the log EL ratio for the mean in a form frequently used in practice and given as

$$lr_e(\theta) = lr(\theta) I\{A_n(\theta)\} + \log(D_n) I\{A_n^c(\theta)\}, \qquad (5)$$

where $I(\cdot)$ is the indicator function, $A_n(\theta) = \left\{\min_{i=1,\ldots,n}(X_i) \leq \theta \leq \max_{i=1,\ldots,n}(X_i)\right\}$ defines the event of $\theta$ when the probability weights $p_i$, $i=1,\ldots,n$, can be computed, $A_n^c(\theta)$ is the complement of the event $A_n(\theta)$ and a specified sequence $D_n$ provides value of $-2\log(D_n)$ to be arbitrarily large (e.g., $D_n \propto \exp(-c_0 n)$, for some constant $c_0$), suggesting to reject the null hypothesis, when testing $EX_1 = \theta$ provided that $I\{A_n^c(\theta)\} = 1$. The definition at (5) is widely applied in practice and we refer the reader, e.g. to the R package *emplik* (R Development Core Team 2015). In the general EL setting (4) the problem of determining the set of $\theta$'s when $p_i$, $i=1,\ldots,n$, can be computed is not a simple task. It requires complicated considerations based on convexifications of sets involved in the EL construction (Owen 2001). Note also that Equation (5)-type forms do not ensure the existence of the expectation $E\{lr_e(\theta)\}$, which is needed in the context of $I_e(\pi)$. For example, in the case of the EL for the mean, when $\theta$ is close to the limits of $A_n(\theta)$, i.e., $\theta \approx \min_{i=1,\ldots,n}(X_i)$ or $\theta \approx \max_{i=1,\ldots,n}(X_i)$, the expectation may be unbounded; see Remark A2 in the SM (Appendix B) for additional details.

Thus we propose to adjust the definition of log EL ratio to be

$$lr_e(\theta) = lr(\theta) I\{B_n(\theta)\} + \log(D_n) I\{B_n^c(\theta)\}, \text{ with } B_n(\theta) = \{w_1(\theta) > M, w_2(\theta) > M\}, \quad (6)$$

where $D_n$ is independent of data $X$, $D_n \to 0$, $w_1 = \sum_{i=1}^n \{G(X_i,\theta)\}^2 I\{G(X_i,\theta) < 0\}$ and $w_2 = \sum_{i=1}^n \{G(X_i,\theta)\}^2 I\{G(X_i,\theta) > 0\}$. We have, for example when $G(X_i,\theta) = X_i - \theta$, $B_n(\theta) = \left[\sum_{i=1}^n (X_i - \theta)^2 I\{(X_i - \theta) < 0\} > M, \sum_{i=1}^n (X_i - \theta)^2 I\{(X_i - \theta) > 0\} > M\right]$. This ensures $\theta \in \{\min_{i=1,\dots,n}(X_i), \max_{i=1,\dots,n}(X_i)\}$ and the existence of $E\{lr_e(\theta)\}$; see Remark A2 in the SM (Appendix B) and the proof schemes of Lemmas A1 and A5 shown in Appendix A for additional details regarding definition (6).

Using the structure at (6), we now denote the expected K-L distance as

$$I_n(\pi) = \int \pi(\theta) E[\log\{\pi_e(\theta|X)/\pi(\theta)\}] d\theta, \quad (7)$$

where $\pi_e(\theta|X) = \pi(\theta) \exp\{lr_e(\theta)\} / \int \pi(\xi) \exp\{lr_e(\xi)\} d\xi$. Consider the EL in (4), requiring that $\partial G(X_i,\theta)/\partial\theta < 0$ (or $\partial G(X_i,\theta)/\partial\theta > 0$), for all $i=1,\dots,n$, and $\pi(\theta)$ is twice continuously differentiable in a neighborhood of the MELE $\hat{\theta}$, where $\hat{\theta}$ is the solution of $\sum_{i=1}^n G(X_i,\hat{\theta}) = 0$. These conditions provide the similarity between the behavior of the EL and the parametric likelihood with respect to the parameter $\theta$; for details see Lemma A1 in Vexler, Ge, and Hutson (2014) as well as Zhong and Ghosh (2016, p. 3019). In a similar manner of restrictions used in Qin and Lawless (1994), we assume that $|\partial G(x,\theta)/d\theta|$ and $|\partial^2 G(x,\theta)/d\theta^2|$ are bounded by some function $Q(x)$ with $E\{Q(X_1)^4\} < \infty$ when $\theta$ belongs to an interval of the MELE $\hat{\theta}$. In this case, we obtain the asymptotic result regarding $I_n(\pi)$.

*Proposition 1.* Assume for $\gamma > 0$, $E|G(X_1,\theta)|^{8+\gamma} < \infty$. Then

$$I_n(\pi) = \frac{1}{2}\log\frac{n}{2\pi e} + \int \pi(\theta) \log\left[\pi(\theta)\{\sigma^2(\theta)\}^{0.5}\right]^{-1} d\theta + o(1), \quad \text{as } n \to \infty \quad (8)$$

where $\sigma^2(\theta) = E\{G(X,\theta)\}^2 / [E\{G'(X,\theta)\}]^2$ and $G'(X,\theta) = \partial G(X,\theta)/\partial\theta$.

This result can be considered as a nonparametric version of the asymptotic conclusion given at (2). Since the integral in (8) is the only term involving the prior density function $\pi(\theta)$, maximizing that integral results in the following corollary related to the EL prior. The proof of Proposition 1, shown in Appendix A, consists of Lemmas A1 and A4 that have an independent interest in evaluations of the EL-type constructions. For example, see Remark 1 in Appendix A.

*Corollary 1.* Let the conditions of Proposition 1 hold. Then the prior density function $\pi(\theta)$, which satisfies $\pi(\theta) \propto \{\sigma^2(\theta)\}^{-0.5}$, maximizes $I_n(\pi)$.

Thus, the prior function $\pi(\theta)$ that satisfies $\pi(\theta) \propto \{\sigma^2(\theta)\}^{-0.5}$ provides asymptotically maximum information distance in the context of $I_n(\pi)$ between the prior and posterior functions based on the EL.

## 3 Bias Reduction of the MELE

In parametric statistics, one can reduce the bias of maximum likelihood estimates using Jeffreys prior in order to penalize the corresponding parametric likelihood functions (Firth 1992, 1993). We propose applying the EL prior in order to reduce the bias associated with the MELE, $\hat{\theta} = \operatorname{argmax}_\theta \{l(\theta)\}$, where $l(\theta)$ is defined in (4). In this case, the estimator $\hat{\theta}$ is the solution of the estimating equation $n^{-1}\sum_{i=1}^n G(X_i, \hat{\theta}) = 0$ (Qin and lawless 1994). Define the penalized MELE $\tilde{\theta} = \arg\max_\theta \left[ \exp\{l(\theta)\} \{\sigma^2(\theta)\}^{-0.5} \right]$, where the EL prior $\pi(\theta)$ is applied following Corollary 1. Newey and Smith (2004) showed the higher order expansion of $\hat{\theta}$ has the form

$$\hat{\theta} = \theta_0 + T_n + o_p(n^{-1}), \ T_n = O_p(n^{-1}),$$

where $\theta_0$ satisfies $E\{G(X_1, \theta_0)\} = 0$. In this context, the first order bias of $\hat{\theta}$ is defined as $Bias(\hat{\theta}) = E(T_n)$. We assume that the identification and regularity conditions presented in Newey and Smith (2004) are satisfied. In particular, these conditions restrict forms of the function $G$ in order to have a unique solution of $\theta$ satisfying the equation $E\{G(X_1,\theta_0)\}=0$ and the ability of the function $E\{G(X_1,\theta)\}$ to be explored using the Taylor theorem. In the following propositions, we present the first order bias forms of the MELE $\hat{\theta}$ and the proposed penalized MELE $\tilde{\theta}$.

*Proposition 2.* The first order bias form of $\hat{\theta}$ satisfies

$$Bias(\hat{\theta}) = \frac{E\{G(X_1,\theta_0)G'(X_1,\theta_0)\}}{n[E\{G'(X_1,\theta_0)\}]^2} - \frac{1}{2}\frac{E\{G(X_1,\theta_0)\}^2 E\{G''(X_1,\theta_0)\}}{n[E\{G'(X_1,\theta_0)\}]^3} + o(n^{-1}),$$

where $G'(X,\theta_0) = \partial G(X,\theta_0)/\partial \theta$ and $G''(X,\theta_0) = \partial^2 G(X,\theta_0)/\partial \theta^2$.

The proof of this proposition can be found in Newey and Smith (2004).

*Proposition 3.* The first order bias form of $\tilde{\theta}$ satisfies

$$Bias(\tilde{\theta}) = \frac{1}{2}\frac{E\{G(X_1,\theta_0)\}^2 E\{G''(X_1,\theta_0)\}}{n[E\{G'(X_1,\theta_0)\}]^3} + o(n^{-1}), \text{ as } n \to \infty.$$

Thus when $E\{G''(X_1,\theta_0)\}=0$, the penalized MELE $\tilde{\theta}$ achieves the first order bias reduction such that the first order bias $Bias(\tilde{\theta})=0$. In scenarios with $|Bias(\tilde{\theta})| \leq |Bias(\hat{\theta})|$, the asymptotic first order bias of $\tilde{\theta}$ is smaller than that of $\hat{\theta}$.

# 4    Simulations

In this section, we evaluate numerically the performance of the proposed bias reduction method using two scenarios related to the estimation of $\theta_0$ that satisfies $E\{G(X_1,\theta_0)\}=0$. These scenarios correspond to the choices of $G(x,\theta)$ as (i) $G(x,\theta) = x^2 - 2\theta x$ and (ii) $G(x,\theta) = \exp(x) - \exp(\mu + \theta/2)$, where $\mu = 0, 1, 1.5$. In order to provide MC examinations of the proposed procedure, we generated 10,000 samples of sizes $n$=15, 25, 50, 75, and 100 from the Normal, Exponential, Chi-squared and Lognormal distributions. We focused on normal and lognormal distributed data, since one can show the EL approach is commonly very efficient in analyzing normally distributed observations, whereas applications of EL methods can be inaccurate when underlying data are skewed (Vexler et al. 2009, 2016; Yu, Vexler, and Tian 2010). In setting (*i*), the true parameter value is $\theta_0 = E(X_1^2)/\{2E(X_1)\}$ for $E(X_1) \neq 0$ and $\theta_0 = 5.2$, *1.0, 1.5, 0.728* in accordance with the underlying distributions $N(10, 2)$, $Exp(1)$, $Chisq(1)$ and $Lognormal(0, 0.5)$, respectively. The corresponding EL prior distribution satisfies $\pi(\theta) \propto \{\sigma^2(\theta)\}^{-1/2}$, where $\sigma^2(\theta)$ has the form $\{4\theta^2 E(X_1^2) - 4\theta E(X_1^3) + E(X_1^4)\}/\{2E(X_1)\}^2$, which is derived using Corollary 1. In this case, the MELE $\hat{\theta}$ and the penalized MELE $\tilde{\theta}$ satisfy $\hat{\theta} = \text{argmax}_\theta \{l(\theta)\}$ and $\tilde{\theta} = \text{argmax}_\theta \{l(\theta)\{\sigma^2(\theta)\}^{-0.5}\}$, where $l(\theta)$ is defined in (4). Since in case (*i*) $G''(X_i, \theta) = 0$, Propositions 2 & 3 show asymptotically that $Bias(\tilde{\theta}) = 0$, whereas $Bias(\hat{\theta}) = \{-2E(X_1^3) + 4\theta E(X_1^2)\}/\{n(2E(X_1))^2\}$.

In setting (*ii*), data points were generated from the $N(\mu, \sigma^2)$ distribution, where $\mu = 0, 1, 1.5$ and $\sigma^2 = 1, 1.5$. In this case, the true parameter value is $\theta_0 = \sigma^2 = 1, 1.5$ and the corresponding EL prior distribution satisfies $\pi(\theta) \propto \{\sigma^2(\theta)\}^{-1/2}$, with $\sigma^2(\theta) = \{\exp(2\mu + 2\theta) - 2\exp(2\mu + \theta) + \exp(\mu + \theta/2)\}/\{\exp(\mu + \theta/2)/2\}^2$. In setting (*ii*), we have $E\{G(X_1, \theta_0)G'(X_1, \theta_0)\} = -0.5\exp(\mu + \theta/2)E\{G(X_1, \theta_0)\} = 0$. Then Propositions 2 & 3 provide asymptotically that $Bias(\tilde{\theta}) = -Bias(\hat{\theta})$. The MC results related to scenarios (*i*) and (*ii*) are presented in Tables 1 and 2, respectively.

Under setting (*i*) with $X_1, ..., X_n \sim N(10, 2)$, $Exp(1)$, $Chisq(1)$ and $Lognormal(0, 0.5)$, when sample sizes of *n*=15 and 25, we show in Table 1 that the penalized MELEs have smaller biases on average than those of the MELEs. Table 1 demonstrates that the mean squared errors (MSE) of the penalized MELEs, $MSE(\tilde{\theta})$, are less than those of the MELEs, $MSE(\hat{\theta})$. Similar results can also be observed for moderate sample sizes of *n*=50 and *n*=75. When sample size increases to *n*=100 and *n*=150, both the MELEs and the penalized MELEs are very close to the true parameter value $\theta_0$ and the MELEs still have less bias on average than those of the MLEs. The MSEs of the MELEs and the penalized MELEs are almost the same. The MC study confirms that in setting (*i*) the penalized MELEs have less bias on average than the MELEs. This result is in accordance with the Propositions 2 and 3 that show asymptotically the first order bias $Bias(\tilde{\theta}) = 0$ and $Bias(\hat{\theta}) \neq 0$ in setting (*i*).

Given setting (*ii*) and in accordance with Propositions 2 and 3 in Section 3, we can expect that the MELEs on average underestimate the true parameter value $\theta_0$ and the penalized MELEs overestimate the true parameter value $\theta_0$ for considered cases since the asymptotically first order biases of $\hat{\theta}$ and $\tilde{\theta}$ have the asymptotical relation $Bias(\tilde{\theta}) = -Bias(\hat{\theta})$. Here, the MC study confirms this conclusion as shown in Table 2. For relatively small sample sizes of *n*=15 and *n*=25, the penalized MELEs still provide estimators with smaller biases on average than those of the MELEs. The mean squared errors (MSE) of the penalized MELEs, $MSE(\tilde{\theta})$, are less than those of the MELEs, $MSE(\hat{\theta})$. For moderate sample sizes of *n*=50 and *n*=75, the penalized MELEs provide smaller biases on average than those of the MELEs. The mean square errors (MSE) of the penalized MELEs, $MSE(\tilde{\theta})$, are a little larger than those of the MELEs, $MSE(\hat{\theta})$. When sample size increases to 100 and 150, both the MELEs and the penalized MELEs are very close to the true parameter value $\theta_0$.

**Table 1**: The Monte Carlo expectations of the MELE $\hat{\theta}$ and the penalized MELE $\tilde{\theta}$ of $\theta$, when $G(X,\theta) = X^2 - 2\theta X$.

| | N(10,2) | | | | | | Exp (1) | | | | | |
|---|---|---|---|---|---|---|---|---|---|---|---|---|
| *n* | 15 | 25 | 50 | 75 | 100 | 150 | 15 | 25 | 50 | 75 | 100 | 150 |
| $\theta_0$ | 5.200 | 5.200 | 5.200 | 5.200 | 5.200 | 5.200 | 1.000 | 1.000 | 1.000 | 1.000 | 1.000 | 1.000 |
| $\hat{\theta}$ | 5.185 | 5.193 | 5.195 | 5.196 | 5.197 | 5.198 | 0.934 | 0.959 | 0.982 | 0.981 | 0.991 | 0.994 |
| $\tilde{\theta}$ | 5.194 | 5.200 | 5.199 | 5.198 | 5.199 | 5.200 | 0.945 | 0.971 | 0.993 | 0.990 | 0.999 | 0.999 |
| $n(\hat{\theta} - \theta_0)$ | -0.229 | -0.167 | -0.233 | -0.312 | -0.271 | -0.251 | -0.984 | -1.015 | -0.882 | -1.403 | -0.878 | -0.907 |

|  | Chisq(1) | | | | | | Lognormal(0,0.5) | | | | | |
|---|---|---|---|---|---|---|---|---|---|---|---|---|
| $n(\tilde{\theta}-\theta_0)$ | -0.092 | -0.014 | -0.062 | -0.133 | -0.090 | -0.066 | -0.818 | -0.720 | -0.357 | -0.743 | -0.148 | -0.088 |
| $MSE(\hat{\theta})$ | 0.067 | 0.041 | 0.020 | 0.014 | 0.010 | 0.007 | 0.104 | 0.070 | 0.039 | 0.024 | 0.020 | 0.013 |
| $MSE(\tilde{\theta})$ | 0.061 | 0.038 | 0.020 | 0.013 | 0.010 | 0.007 | 0.095 | 0.065 | 0.035 | 0.021 | 0.020 | 0.013 |

|  | Chisq(1) | | | | | | Lognormal(0,0.5) | | | | | |
|---|---|---|---|---|---|---|---|---|---|---|---|---|
| $n$ | 15 | 25 | 50 | 75 | 100 | 150 | 15 | 25 | 50 | 75 | 100 | 150 |
| $\theta_0$ | 1.500 | 1.500 | 1.500 | 1.500 | 1.500 | 1.500 | 0.728 | 0.728 | 0.728 | 0.728 | 0.728 | 0.727 |
| $\hat{\theta}$ | 1.305 | 1.389 | 1.453 | 1.458 | 1.473 | 1.480 | 0.711 | 0.718 | 0.722 | 0.723 | 0.725 | 0.726 |
| $\tilde{\theta}$ | 1.317 | 1.408 | 1.473 | 1.477 | 1.490 | 1.494 | 0.715 | 0.721 | 0.725 | 0.726 | 0.727 | 0.727 |
| $n(\hat{\theta}-\theta_0)$ | -2.921 | -2.763 | -2.330 | -3.139 | -2.701 | -2.980 | -0.243 | -0.240 | -0.279 | -0.314 | -0.293 | -0.287 |
| $n(\tilde{\theta}-\theta_0)$ | -2.741 | -2.304 | -1.352 | -1.749 | -1.014 | -0.876 | -0.192 | -0.153 | -0.135 | -0.139 | -0.098 | -0.071 |
| $MSE(\hat{\theta})$ | 0.455 | 0.313 | 0.197 | 0.133 | 0.101 | 0.068 | 0.020 | 0.013 | 0.007 | 0.005 | 0.004 | 0.002 |
| $MSE(\tilde{\theta})$ | 0.405 | 0.283 | 0.187 | 0.132 | 0.102 | 0.070 | 0.007 | 0.005 | 0.004 | 0.003 | 0.002 | 0.002 |

**Table 2**: The Monte Carlo expectations of the MELE $\hat{\theta}$ and penalized MELE $\tilde{\theta}$ of $\theta$, when $G(X,\theta) = \exp(X) - \exp(\mu + \theta/2)$.

|  | N(0,1) | | | | | | N(1,1) | | | | | |
|---|---|---|---|---|---|---|---|---|---|---|---|---|
| $n$ | 15 | 25 | 50 | 75 | 100 | 150 | 15 | 25 | 50 | 75 | 100 | 150 |
| $\theta_0$ | 1.000 | 1.000 | 1.000 | 1.000 | 1.000 | 1.000 | 1.000 | 1.000 | 1.000 | 1.000 | 1.000 | 1.000 |
| $\hat{\theta}$ | 0.893 | 0.937 | 0.967 | 0.978 | 0.987 | 0.988 | 0.898 | 0.936 | 0.968 | 0.972 | 0.984 | 0.989 |
| $\tilde{\theta}$ | 1.029 | 1.030 | 1.022 | 1.017 | 1.017 | 1.009 | 1.032 | 1.029 | 1.023 | 1.011 | 1.015 | 1.011 |
| $n(\hat{\theta}-\theta_0)$ | -1.612 | -1.587 | -1.642 | -1.628 | -1.337 | -1.817 | -1.533 | -1.608 | -1.603 | -2.115 | -1.573 | -1.596 |
| $n(\tilde{\theta}-\theta_0)$ | 0.428 | 0.745 | 1.087 | 1.292 | 1.698 | 1.329 | 0.474 | 0.723 | 1.156 | 0.842 | 1.462 | 1.581 |
| $MSE(\hat{\theta})$ | 0.406 | 0.257 | 0.133 | 0.087 | 0.067 | 0.044 | 0.403 | 0.249 | 0.131 | 0.087 | 0.066 | 0.045 |
| $MSE(\tilde{\theta})$ | 0.367 | 0.257 | 0.134 | 0.088 | 0.069 | 0.045 | 0.370 | 0.247 | 0.132 | 0.088 | 0.067 | 0.046 |

|  | N(1.5,1) | | | | | | N(1.5,1.5) | | | | | |
|---|---|---|---|---|---|---|---|---|---|---|---|---|
| $n$ | 15 | 25 | 50 | 75 | 100 | 150 | 15 | 25 | 50 | 75 | 100 | 150 |

| $\theta_0$ | 1.000 | 1.000 | 1.000 | 1.000 | 1.000 | 1.000 | 1.500 | 1.500 | 1.500 | 1.500 | 1.500 | 1.500 |
|---|---|---|---|---|---|---|---|---|---|---|---|---|
| $\hat{\theta}$ | 0.887 | 0.934 | 0.971 | 0.977 | 0.985 | 0.990 | 1.324 | 1.386 | 1.440 | 1.459 | 1.464 | 1.475 |
| $\tilde{\theta}$ | 1.022 | 1.027 | 1.025 | 1.016 | 1.015 | 1.011 | 1.518 | 1.531 | 1.532 | 1.528 | 1.519 | 1.514 |
| $n(\hat{\theta}-\theta_0)$ | -1.696 | -1.660 | -1.466 | -1.749 | -1.507 | -1.454 | -2.641 | -2.842 | -3.024 | -3.043 | -3.591 | -3.679 |
| $n(\tilde{\theta}-\theta_0)$ | 0.331 | 0.664 | 1.264 | 1.173 | 1.543 | 1.711 | 0.264 | 0.777 | 1.622 | 2.108 | 1.872 | 2.147 |
| $MSE(\hat{\theta})$ | 0.401 | 0.260 | 0.130 | 0.089 | 0.067 | 0.044 | 0.732 | 0.455 | 0.245 | 0.173 | 0.125 | 0.087 |
| $MSE(\tilde{\theta})$ | 0.363 | 0.257 | 0.134 | 0.090 | 0.070 | 0.046 | 0.706 | 0.473 | 0.266 | 0.180 | 0.136 | 0.094 |

## 5   Real Data Example

In this section, a real data example is presented in order to illustrate the practical applicability of the proposed bias reduction method. The example is based on data from a study that evaluated the association between biomarkers and myocardial infarction (MI). The study was focused on the residents of Erie and Niagara counties, 35-79 years of age (Schisterman et al. 2001). The New York State department of Motor Vehicles drivers' license rolls was used as the sampling frame for adults between the age of 35 and 65 years, while the elderly sample (age 65-79) was randomly chosen from the Health Care Financing Administration database. We consider the biomarker "Vitamin E" supplement that is often used to quantify antioxidant status of an individual and could prevent heart disease (Rimm et al. 1993). A total of 2390 measurements of Vitamin E were evaluated by the study. 547 of them were collected from cases who survived on MI and the other 1843 from controls who had no previous MI. We denote the data points in control and case groups as ($Y_1,\ldots, Y_{1843}$) and ($X_1,\ldots, X_{547}$), respectively.

We applied the proposed bias reduction method in the context of the *M*-estimators of $\theta_Y$ and $\theta_X$ that are given as the roots of $E(Y_1-\theta_Y)^3 = 0$ and $E(X_1-\theta_X)^3 = 0$, respectively. In this case, the corresponding MELEs, $\hat{\theta}_Y$ and $\hat{\theta}_X$, are solutions of $1843^{-1}\sum_{i=1}^{1843}G(Y_i,\hat{\theta}_Y)=0$ and $547^{-1}\sum_{i=1}^{547}G(X_i,\hat{\theta}_X)=0$ where $G(u,\theta)=(u-\theta)^3$. We illustrate the proposed method employing the following bootstrap-type algorithm: The strategy is that samples of sizes $n=$ 25, 50, 75 and 100 are randomly selected from the Vitamin E measurements related to control and case groups, respectively. These samples are then used to compute values of the sample MELEs, $\hat{\theta}_{Yn}$ and $\hat{\theta}_{Xn}$, and the sample penalized MELEs, $\tilde{\theta}_{Yn}$ and $\tilde{\theta}_{Xn}$, defined in Section 3 for the control and case groups, respectively. In order to obtain penalized MELEs, the corresponding EL priors satisfies $\pi_Y(\theta) \propto \{\sigma_Y^2(\theta)\}^{-1/2}$ with

$$\sigma_Y^2(\theta) = E\{G(Y_1,\theta)\}^2 / [E\{G'(Y_1,\theta)\}]^2 \quad \text{and} \quad \pi_X(\theta) \propto \{\sigma_X^2(\theta)\}^{-1/2} \quad \text{with}$$

$\sigma_X^2(\theta) = E\{G(X_1,\theta)\}^2 / [E\{G'(X_1,\theta)\}]^2$ for control and case groups, respectively. Since the underlying data distributions are unknown, we can use the sample moments estimators of $E(X_1), E(X_1^2), E(X_1^3)$, $E(Y_1), E(Y_1^2)$ and $E(Y_1^3)$ in order to approximate the prior distributions. The evaluated MELEs, $\hat{\theta}_{Yn}$ and $\hat{\theta}_{Xn}$, and penalized MELEs, $\tilde{\theta}_{Yn}$ and $\tilde{\theta}_{Xn}$, were calculated based on samples of sizes $n=$ 25, 50, 75 and 100 that are smaller than the rest of data (1843-$n$) and (547-$n$) (the amount of observations that are not used to compute evaluated estimators with respect to case and control groups). Then the rest of the data points, 1843-$n$ and 547-$n$, are used to estimate the true parameter values $\theta_{Yn}$ and $\theta_{Xn}$ that satisfy $\sum_{i=1}^{1843-n}(Y_i - \theta_{Yn})^3 = 0$ and $\sum_{i=1}^{1843-n}(X_i - \theta_{Xn})^3 = 0$ for control group and case group, respectively. In this bootstrap-type algorithm, the R functions *optimize* (R Development Core Team 2015) can be used to compute values of $\hat{\theta}_{Yn}$ and $\tilde{\theta}_{Yn}$ for the control group ($\hat{\theta}_{Xn}$ and $\tilde{\theta}_{Xn}$ for case group). The values of (1843-$n$) and (547-$n$) were chosen to be relatively large so that the calculated estimators $\theta_{Yn}$ and $\theta_{Xn}$ are close to the true theoretical values. The strategy above was repeated 10,000 times in a bootstrap manner to compute the MC average bias of the MELEs and the penalized MELEs. Table 3 presents the results of this data example related to MI. This bootstrap-type test shows that the proposed penalized MELEs on average have less bias than those of the corresponding MELEs for both the control and case groups.

**Table 3**. The bootstrap-type expectations of the MELEs ($\hat{\theta}_{Yn}$ and $\hat{\theta}_{Xn}$) and the penalized MELEs ($\tilde{\theta}_{Yn}$ and $\tilde{\theta}_{Xn}$) for case and control groups

| Control | n=25 | n=50 | n=75 | n=100 |
|---|---|---|---|---|
| $\theta_{Yn}$ | 17.813 | 17.815 | 17.817 | 17.819 |
| $\tilde{\theta}_{Yn} - \theta_{Yn}$ | -0.783 | -0.412 | -0.274 | -0.211 |
| $\hat{\theta}_{Yn} - \theta_{Yn}$ | -1.166 | -0.772 | -0.601 | -0.509 |
| Case | n=25 | n=50 | n=75 | n=100 |
| $\theta_{Xn}$ | 18.558 | 18.553 | 18.534 | 18.517 |
| $\tilde{\theta}_{Xn} - \theta_{Xn}$ | -2.074 | -1.316 | -0.776 | -0.441 |
| $\hat{\theta}_{Xn} - \theta_{Xn}$ | -2.503 | -1.824 | -1.353 | -1.053 |

# 6 Conclusion

In this article, we derived a Jeffreys-type EL prior, focusing on an integrated K-L distance between the EL-based posterior and prior density functions. Rigorous evaluations of this K-L distance are shown to provide a nonparametric counterpart to the classical result obtained in parametric based statistics. We applied the proposed EL prior to develop a methodology for removing the first-order term from the asymptotic bias of solutions of general estimating equations and *M*-estimation schemes. An extensive MC study and a data example confirmed the efficiency and applicability of the proposed method.

## Acknowledgments

Drs. Vexler's effort was supported by the National Institutes of Health (NIH) grant 1G13LM012241-01.

## Appendix A. Proof Schemes

*Proof of Proposition 1*: To prove Proposition 1, we represent Equation (7) as

$$I_n(\pi) = \int \pi(\theta) E\{lr_e(\theta)\} d\theta - \int \pi(\theta) E\left\{\log \int \pi(\xi) \exp\{lr_e(\xi)\} d\xi\right\} d\theta, \quad (A.1)$$

where $lr_e(\theta)$ is defined in (6). The proof scheme of Proposition 1 is based on the following two stages: I) we will evaluate the first component of the right side in (A.1) to show that the expectation $E\{lr_e(\theta)\}$ converges to -1/2, asymptotically. Intuitively this fact follows from the Wilks-type result, $-2lr(\theta) \to \chi_1^2$ as $n \to \infty$ (Owen 2001). However the convergence in expectation requires more complicated considerations; II) regarding the second component of the right side in (A.1), we will use the results related to the marginal EL that are obtained in Vexler, Ge, and Hutson (2014). These results in the EL setting adapt and extend Laplace's method used in the parametric Bayesian analysis (Tierney and Kadane 1986). Technical proofs of the lemmas used in this appendix are presented in the Supplementary Material, Appendix B.

Attending to the first stage of the proof scheme mentioned above, we note that $-2E\{lr_e(\theta)\} = -2E[lr(\theta)I\{B_n(\theta)\}] - 2\log(D_n)\Pr\{B_n^c(\theta)\}$. It is clear that $\log(D_n)\Pr\{B_n^c\} = o(1)$ as $n \to \infty$, since the Chebyshev inequality provides

$$\Pr\left[\sum_{i=1}^n G(X_i, \theta)^2 I\{G(X_i, \theta) < 0\} < M\right] \le E\left|\sum_{i=1}^n [q - G(X_i, \theta)^2 I\{G(X_i, \theta) < 0\}]\right|^3 (nq - M)^{-3}$$

$= O(n^{-1.5})$ with $q = E[\{G(X_1, \theta)\}^2 I\{G(X_1, \theta) < 0\}]$ assuming $E|\{G(X_1, \theta)\}^2 I\{G(X_1, \theta) < 0\}|^3 < \infty$. Next we evaluate the main term $-2E[lr(\theta)I\{B_n(\theta)\}]$, showing that $-2E[lr(\theta)I\{B_n(\theta)\}] \to 1$ asymptotically. To this end, we rewrite $-2E[lr(\theta)I\{B_n(\theta)\}]$ using two integrals

$$-2E[lr(\theta)I\{B_n(\theta)\}] = \int_0^{b_n} \Pr\{-2lr(\theta) > x, B_n(\theta)\}dx + \int_{b_n}^\infty \Pr\{-2lr(\theta) > x, B_n(\theta)\}dx, \quad (A.2)$$

where $lr(\theta) < 0$, $b_n = n^{1-\varepsilon}$ and $\varepsilon > 0$. The following lemma considers the first component of the right side of (A.2).

*Lemma A1*. Assume that, for $\gamma > 0$, $E|G(X_1, \theta)|^{4+\gamma} < \infty$. Then, defining $b_n = n^{1-\varepsilon}$ with $0 < \varepsilon < 1$, we have

$$\int_0^{b_n} \Pr\{-2lr(\theta) > x, B_n(\theta)\}dx \to 1, \text{ as } n \to \infty.$$

The corresponding lemma A1 proof scheme is based on results shown in DiCiccio, Hall, and Romano (1991).

In order to show that the remainder term $\int_{b_n}^{\infty} \Pr\{-2lr(X|\theta) > x, B_n(\theta)\}dx$ in (A.2) vanishes to zero, we present the lemmas below. These lemmas incorporate the Lagrange multiplier $\lambda$ related to the Lagrangian equation

$$H = \sum_{i=1}^{n} \log p_i + \lambda_1\left(1 - \sum_{i=1}^{n} p_i\right) - \lambda \sum_{i=1}^{n} p_i G(X_i, \theta),$$

in order to maximize the EL, $\prod_{i=1}^{n} p_i$, given the constraints $\sum_{i=1}^{n} p_i = 1$ and $\sum_{i=1}^{n} p_i G(X_i, \theta) = 0$. It can be easily shown that $\lambda_1 = n$, $p_i = \{n + \lambda G(X_i, \theta)\}^{-1}$, $i=1,...,n$, and $\lambda$ is the solution of $n^{-1}\sum_{i=1}^{n} G(X_i, \theta)/\{n + \lambda G(X_i, \theta)\} = 0$; for details, see Qin and Lawless (1994).

*Lemma A2.* We have that $\lambda \geq 0$ if and only if $\sum_{i=1}^{n} G(X_i, \theta) \geq 0$; $\lambda < 0$ if and only if $\sum_{i=1}^{n} G(X_i, \theta) < 0$.

*Lemma A3.* The Lagrange multiplier $\lambda$ satisfies $\lambda = \sum_{i=1}^{n} G(X_i, \theta)/\sum_{i=1}^{n} \{G(X_i, \theta)\}^2 p_i$.

*Lemma A4.* If $\lambda \geq 0$, we have $0 \leq \lambda < n\sum_{i=1}^{n} G(X_i, \theta)\left[\sum_{i=1}^{n} G(X_i, \theta)^2 I\{G(X_i, \theta) < 0\}\right]^{-1}$; if $\lambda \leq 0$, we have

$$n\sum_{i=1}^{n} G(X_i, \theta)\left[\sum_{i=1}^{n} G(X_i, \theta)^2 I\{G(X_i, \theta) > 0\}\right]^{-1} < \lambda < 0.$$

Remark 1. Lemma A4 provides the exact non-asymptotic bounds for $\lambda$. Owen (1988) used very complicated considerations to obtain the approximate bounds for $\lambda$ as $n \to \infty$. Lemma A4 immediately demonstrates that $\lambda = O_p(n^{1/2})$ when $E\{G(X_1, \theta)\} = 0$. Lemmas A1 & A4 can be useful in the context of numerical computations of ELs, providing, e.g., the exact bounds for $\lambda$ that is a numerical solution of $n^{-1}\sum_{i=1}^{n} G(X_i, \theta)/\{n + \lambda G(X_i, \theta)\} = 0$.

By virtue of Lemmas A1-A4, we conclude that the remainder term asymptotically vanishes with the following lemma.

*Lemma A5.* Assume that for $\gamma > 0$, $E|G(X_1,\theta)|^{8+\gamma} < \infty$. Then

$$\int_{b_n}^{\infty} \Pr\{-2lr(X \mid \theta) > x, B_n(\theta)\}dx = o(1), \text{ as } n \to \infty.$$

Thus we show that the remainder term $\int_{b_n}^{\infty} \Pr\{-2lr(X \mid \theta) > x, B_n(\theta)\}dx$ in (A.2) vanishes asymptotically to zero. Then by virtue of (A.1), we have $-2E\{lr_e(\theta)\} \to 1$. This completes the first stage of the proof scheme of Proposition 1.

In the second stage of the proof scheme of Proposition 1, we analyze the term $\int \pi(\theta)E\left[\log \int \pi(\zeta)\exp\{lr_e(\zeta)\}d\zeta\right]d\theta$ at (A.1). In a similar manner to the evaluation of $E\{lr_e(\theta)\}$ shown above, we prove that

$$E\left[\log \int \pi(\zeta)\exp\{lr_e(\zeta)\}d\zeta\right] = \log\left[\pi(\theta)\{2\pi\sigma^2(\theta)/n\}^{1/2}\right] + o(1),$$

where $\sigma^2(\theta) = E[G(X,\theta)]^2 /[EG'(X,\theta)]^2$ with $G'(X,\theta) = dG(X,\theta)/d\theta$. Intuitively this result follows from the fact that $\int \pi(\zeta)\exp\{lr(\zeta)\}d\zeta = \pi(\theta_M)(2\pi\sigma_M^2/n)^{1/2} + o_p(n^{-1/2})$, where $\sigma_M^2 = n^{-1}\sum_{i=1}^{n}\{G(X_i,\hat\theta)\}^2 \left\{n^{-1}\sum_{i=1}^{n}\partial G(X_i,\hat\theta)/\partial\theta\right\}^{-2}$ and $\hat\theta$ is the solution of $\sum_{i=1}^{n} G(X_i,\hat\theta) = 0$ (Vexler, Ge, and Hutson 2014). Following the algorithm of proofs shown in Vexler, Ge, and Hutson (2014), we denote a positive sequence $\varphi_n = n^{1/6-\beta} \to \infty$ with the property that $n^{-1/2}\varphi_n \to 0$ where $\beta \in (0,1/6)$ and focus on the following expression

$$\int \exp\{lr_e(\zeta)\}\pi(\zeta)d\zeta = \int_{\hat\theta-\varphi_n n^{-1/2}}^{\hat\theta+\varphi_n n^{-1/2}} \exp\{lr_e(\zeta)\}\pi(\zeta)d\zeta + R_n,$$

where the remainder term $R_n = \int_{-\infty}^{\hat\theta-\varphi_n n^{-1/2}} \exp\{lr_e(\zeta)\}\pi(\zeta)d\zeta + \int_{\hat\theta+\varphi_n n^{-1/2}}^{\infty} \exp\{lr_e(\zeta)\}\pi(\zeta)d\zeta$.

By virtue of that $\log\left[\int \exp\{lr_e(\zeta)\}\pi(\zeta)d\zeta\right] < 0$, we have the following two inequalities

$$E\left[\log \int \exp\{lr_e(\zeta)\}\pi(\zeta)d\zeta\right]$$

$$\leq E\left[\left\{\log\left(\int_{\hat{\theta}-\varphi_n n^{-1/2}}^{\hat{\theta}+\varphi_n n^{-1/2}} \exp\{lr_e(\zeta)\}\pi(\zeta)d\zeta + R_n\right)\right\}I(R_n < d_n)\right],$$

$$\leq E\left[\log\left\{\int_{\hat{\theta}-\varphi_n n^{-1/2}}^{\hat{\theta}+\varphi_n n^{-1/2}} \exp\{lr_e(\zeta)\}\pi(\zeta)d\zeta + d_n\right\}\right] - E\left[\left\{\log\left(\int_{\hat{\theta}-\varphi_n n^{-1/2}}^{\hat{\theta}+\varphi_n n^{-1/2}} \exp\{lr_e(\zeta)\}\pi(\zeta)d\zeta + d_n\right)\right\}I(R_n \geq d_n)\right], \text{and}$$

$$E\left[\log\left\{\int \exp\{lr_e(\zeta)\}\pi(\zeta)d\zeta\right\}\right] \geq E\left[\log\left\{\int_{\hat{\theta}-\varphi_n n^{-1/2}}^{\hat{\theta}+\varphi_n n^{-1/2}} \exp\{lr_e(\zeta)\}\pi(\zeta)d\zeta\right\}\right], \tag{A.3}$$

where we use $d_n = \exp(-n^\tau)$ with $0 < \tau < 1/3 - 2\beta$. We will show that the upper and lower bounds in (A.3) converge to a same value as $n \to \infty$. Towards this end, we derive a bound for $-\log\left\{\int_{\hat{\theta}-\varphi_n n^{-1/2}}^{\hat{\theta}+\varphi_n n^{-1/2}} \exp\{lr_e(\zeta)\}\pi(\zeta)d\zeta\right\}$ in the next lemma. This bound will assist in evaluating the remainder term $E\left[\log\left\{\int_{\hat{\theta}-\varphi_n n^{-1/2}}^{\hat{\theta}+\varphi_n n^{-1/2}} \exp\{lr_e(\zeta)\}\pi(\zeta)d\zeta + d_n\right\}I(R_n \geq d_n)\right]$.

*Lemma A6.* Assume that $|\partial G(x,\theta)/\partial \theta|$ is bounded by some function $Q(x)$ for $\theta \in [\hat{\theta} - \varphi_n n^{-1/2}, \hat{\theta} + \varphi_n n^{-1/2}]$ where $\varphi_n = n^{1/6-\beta}$ with $\beta \in (0, 1/6)$. Then

$$0 < -\log\left\{\int_{\hat{\theta}-\varphi_n n^{-1/2}}^{\hat{\theta}+\varphi_n n^{-1/2}} e^{lr_e(\zeta)}\pi(\zeta)d\zeta\right\} \leq \left\{2\varphi_n^2 n\left(\sum_{i=1}^n Q(X_i)\right)^2 / M - \log(D_n)\right\} - \log\left\{\int_{\hat{\theta}-\varphi_n n^{-1/2}}^{\hat{\theta}+\varphi_n n^{-1/2}} \pi(\zeta)d\zeta\right\},$$

where $D_n = \exp(-c_0 n)$ and $c_0$ is a positive constant.

*Lemma A7.* Assume that $|\partial G(x,\theta)/\partial \theta|$ is bounded by some function $Q(x)$ with $E\{Q(X_1)^4\} < \infty$ for $\theta \in [\hat{\theta} - \varphi_n n^{-1/2}, \hat{\theta} + \varphi_n n^{-1/2}]$ where $\varphi_n = n^{1/6-\beta}$ with $\beta \in (0, 1/6)$. Then

$$-E\left[\log\left\{\int_{\hat{\theta}-\varphi_n n^{-1/2}}^{\hat{\theta}+\varphi_n n^{-1/2}} \exp\{lr_e(\zeta)\}\pi(\zeta)d\zeta + d_n\right\}I(R_n \geq d_n)\right] \to 0,$$

where $R_n = \int_{-\infty}^{\hat{\theta}-\varphi_n n^{-1/2}} \exp\{lr_e(\zeta)\}\pi(\zeta)d\zeta + \int_{\hat{\theta}+\varphi_n n^{-1/2}}^{\infty} \exp\{lr_e(\zeta)\}\pi(\zeta)d\zeta$ and $d_n = \exp(-n^{\xi})$ with

$0 < \xi < 1/3 - 2\beta$.

By virtue of Lemmas A6 and A7, using (A.3) we have

$$-E\left[\log\left\{\int_{\hat{\theta}-\varphi_n n^{-1/2}}^{\hat{\theta}+\varphi_n n^{-1/2}} \exp\{lr_e(\zeta)\}\pi(\zeta)d\zeta + d_n\right\}\right] + o(1) \leq -E\left[\log\left\{\int \exp\{lr_e(\zeta)\}\pi(\zeta)d\zeta\right\}\right]$$

$$\leq -E\left[\log\left\{\int_{\hat{\theta}-\varphi_n n^{-1/2}}^{\hat{\theta}+\varphi_n n^{-1/2}} \exp\{lr_e(\zeta)\}\pi(\zeta)d\zeta\right\}\right]. \tag{A.4}$$

Taking into account of (A.4) and the following lemma result, we complete the second stage of the proof of Proposition 1.

*Lemma A8.* Assume that, for some $\gamma > 0$ and all $\theta$, $E|G(X_1,\theta)|^{8+\gamma} < \infty$ and the following conditions are satisfied: $|\partial G(x,\theta)/\partial\theta|$ and $|\partial^2 G(x,\theta)/\partial\theta^2|$ are bounded by some function $Q(x)$ with $E\{Q(X_1)^4\} < \infty$, for $\theta \in [\hat{\theta}-\varphi_n n^{-1/2}, \hat{\theta}+\varphi_n n^{-1/2}]$, where $\varphi_n = n^{1/6-\beta}$, $\beta \in (0,1/6)$. Define $\pi(\theta)$ to be a twice continuously differentiable on $\theta \in [\hat{\theta}-\varphi_n n^{-1/2}, \hat{\theta}+\varphi_n n^{-1/2}]$ prior density function. Then

$$-E\left[\log\left\{\int \pi(\zeta)\exp\{lr_e(\zeta)\}d\zeta\right\}\right] = -\log\left[\pi(\theta)\{2\pi\sigma^2(\theta)/n\}^{1/2}\right] + o(1),$$

where $\sigma^2(\theta) = E\{G(X,\theta)\}^2/\{EG'(X,\theta)\}^2$ and $G'(X,\theta) = \partial G(X,\theta)/\partial\theta$.

The corresponding lemma A8 proof scheme is based on results shown in Vexler, Ge, and Hutson (2014) and Zhong and Ghosh (2016).

Thus lemmas (A.4) and (A.8) complete the proof of Proposition 1.

*Proof of Corollary 1*: The proof of Corollary 1 is technically straightforward and similar to that shown in Lehmann and Casella (1998, pp. 261-262) and thus is omitted.

*Proof of Proposition* 3: Define $H(\theta) = \log\{\exp lr(\theta)\pi(\theta)\} = lr(\theta) + \log\{\pi(\theta)\}$. By the definition of $\tilde{\theta}$ in section 3, it satisfies

$$\partial H(\theta)/\partial\theta|_{\theta=\tilde{\theta}} = 0. \tag{A.5}$$

Since $\pi(\theta) \propto \{\sigma^2(\theta)\}^{-0.5}$, one can directly set the prior function to be $\pi(\theta) = \{\sigma^2(\theta)\}^{-0.5}$ as the penalized function in $H(\theta)$. Denote $A(\theta) = \sigma^2(\theta)$, $A'(\theta) = \partial A(\theta)/\partial\theta$, $A''(\theta) = \partial^2 A(\theta)/\partial\theta^2$, $lr'(\theta) = \partial lr(\theta)/\partial\theta$ and $lr''(\theta) = \partial^2 lr(\theta)/\partial\theta^2$. By the first order Taylor expansion to $\partial H(\tilde{\theta})/\partial\theta = 0$ in (A.5) around $\hat{\theta}$, we obtain

$$dH(\theta)/d\theta|_{\theta=\tilde{\theta}} = 0 = lr'(\hat{\theta}) - \frac{A'(\hat{\theta})}{2A(\hat{\theta})} + \left[lr''(\theta) - \frac{A(\theta)A''(\theta) - \{A'(\theta)\}^2}{2A^2(\theta)}\right]_{\theta=\theta^{**}}(\tilde{\theta} - \hat{\theta}), \tag{A.6}$$

where $\theta^{**} = \hat{\theta} + \rho(\tilde{\theta} - \hat{\theta})$ with $\rho \in (0,1)$.

By virtue of the results that $lr'(\hat{\theta}) = 0$ and $lr''(\theta) = O_p(n)$ (for details see the proof of Lemma A8 in SM, Appendix B), we have the following lemma.

*Lemma A9.* We have

$$\tilde{\theta} - \hat{\theta} = O_p(n^{-1}). \tag{A.7}$$

Then by a second order Taylor expansion to $\partial H(\tilde{\theta})/\partial\theta = 0$ in (A.5) around $\tilde{\theta} = \hat{\theta}$, we have

$$lr'(\hat{\theta}) - \frac{A'(\hat{\theta})}{2A(\hat{\theta})} + lr''(\hat{\theta})(\tilde{\theta} - \hat{\theta}) - \frac{\partial\{A'(\theta)/A(\theta)\}}{2\partial\theta}\bigg|_{\theta=\hat{\theta}}(\tilde{\theta} - \hat{\theta}) + \frac{\partial^3 H(\theta^*)}{\partial\theta^3}(\tilde{\theta} - \hat{\theta})^2 = 0, \tag{A.8}$$

where $\theta^* = \hat{\theta} + \rho_1(\tilde{\theta} - \hat{\theta})$ with $\rho_1 \in (0,1)$.

The above expansion (A.8) and Lemma A9 imply that

$$\tilde{\theta} - \hat{\theta} = \frac{A'(\theta)}{2A(\theta)} \{lr''(\theta)\}^{-1} \bigg|_{\theta=\hat{\theta}} + O_p(n^{-2}). \tag{A.9}$$

The expression of $lr''(\theta)$ can be easily found by taking the derivative of the constraint equation $\sum_{i=1}^{n} G(X_i, \theta)\{n + \lambda G(X_i, \theta)\}^{-1} = 0$ with respect to $\theta$. Then one can obtain the following results regarding (A.9): $E\{lr''(\hat{\theta})\} + nE\{G(X_1, \theta_0)\}^2 / [E\{G'(X_1, \theta_0)\}]^2 \to 0$, $A'(\hat{\theta}) \to A'(\theta_0)$, where

$A'(\theta) = 2E\{G(X_1, \theta)G'(X_1, \theta)\}[E\{G'(X_1, \theta)\}]^{-2} - 2E\{G(X_1, \theta)\}^2 E\{G''(X_1, \theta)\}[E\{G'(X_1, \theta)\}]^{-3}$ (for details see the proof of Lemma A8 in SM, Appendix B).

Thus, it is clear that by (A.9) and Proposition 2 we complete the proof of Proposition 3.

## Appendix B. Supplementary Material

The supplementary material contains: Details of the technical derivations and proofs corresponding to the theoretical results and Lemmas presented in this paper as well as detailed remarks of the relevant research article, Clarke and Yuan (2010).

## References

Bernardo, J. M. 1979. Reference posterior densities for Bayesian inference (with discussion). *Journal of the Royal Statistical Society,* Series B 41, 113-147.

Berger, J. O., Bernardo, J. M., and Sun, D. 2009. The formal definition of reference priors, *The Annals of Statistics,* 37, 905-938.

Chaudhuri, S., and Ghosh, M. 2011. Empirical likelihood for small area estimation. *Biometrika,* 98, 473-480.

Clarke, B. and Yuan, A. 2010. Reference priors for empirical likelihoods, In Chen, M. H., Mueller, P., Sun, D., Ye, K. & Dey, D. K. (eds.). *Frontiers of Statistical Decision Making and Bayesian Analysis: In honor of James O. Berger.* 56–68. New York: Springer.

Daniels, M., and Hogan, J. 2007. *Missing data in longitudinal studies: strategies for Bayesian modeling and sensitivity analysis*, New York: Chapman and Hall.

DiCiccio, T., Hall, P., and Romano, J. 1991. Empirical likelihood is Bartlett-correctable, *The Annals of Statistics,* 19, 1053-1061.

Firth, D. 1992. *Bias reduction, the Jeffreys prior and GLIM*. In L. Fahrmeir, B. Francis, R. Gilchrist & G. Tutz (Eds.). *Advances in GLIM and Statistical Modelling: Proceedings of the GLIM 92 Conference*, Munich, New York, pp. 91–100. Springer.

Firth, D. 1993. Bias reduction of maximum likelihood estimates, *Biometrika,* 80, 27–38.

Hansen, L. P. 1982. Large sample properties of generalized method of moments estimators, *Econometrica,* 50, 1029-1054.

Hartigan, J. A. 1964. Invariant prior distributions, *The Annals of Mathematical Statistics,* 35, 836-845.

Hartigan, J. A. 1998. The maximum likelihood prior, *The Annals of Statistics,* 26, 2083-2103.

Jeffreys, H. 1946. An invariant form for the prior probability in estimation problems, *Proceedings of the Royal Society of London, Series A*, 186, 453-461.

Qin, J., and Lawless, J. 1994. Empirical Likelihood and General Estimating Equations, *The Annals of Statistics,* 22, 300-325.

Lazar, N. A. 2003. Bayesian Empirical Likelihood, *Biometrika,* 90, 319-326.

Lehmann, E. L., and Casella, G. 1998. *Theory of Point Estimation*. New York: John Wiley.

Lindley, D. V. 1956. On a measure of the information provided by an experiment, *The Annals of Mathematical Statistics,* 27, 986-1005.

Newey, W. K., and Smith, R. J. 2004. Higher Order Properties of GMM and Generalized Empirical Likelihood Estimators, *Econometrica,* 72, 219-255.

Owen, A. B. 1988. Empirical likelihood ratio confidence intervals for a single functional, *Biometrika,* 75, 237-49.

Owen, A. 2001. *Empirical Likelihood*, New York: Chapman and Hall.

Rimm E.B., Stampfer M.J., Ascherio A., Giovannucci E., Colditz G.A., and Willett W.C. 1993. Vitamin E Consumption and the Risk of Coronary Heart Disease in Men, *The New England Journal of Medicine*, 328, 1450-1456.

R development core team 2015. R: A Language and Environment for Statistical Computing. Vienna, Austria: R Foundation for Statistical Computing. ISBN 3-900051-07-0, http://www.R-project.org.

Schisterman, E. F., Faraggi, D., Browne, R., Freudenheim, J., Dorn, J., Muti, P., Armstrong, D., Reiser, B., and Trevisan, M. 2001. Tbars and cardiovascular disease in a population-based sample, *Journal of cardiovascular risk,* 8, 219–225.

Tierney, L., and Kadane, J.B. 1986. Accurate approximations for posterior moments and marginal densities, *Journal of the American Statistical Association,* 81, 82-86.

Vexler, A., Liu, S. L., Kang, L., and Hutson, A. D. 2009. Modifications of the Empirical Likelihood Interval Estimation with Improved Coverage Probabilities, Communications in Statistics, Simulation and Computation, 38, 2171-2183.

Vexler, A., Tao, G., and Hutson, A. D. 2014. Posterior expectation based on empirical likelihoods, *Biometrika,* 101, 711-718.

Vexler, A., Zou, L., and Hutson, A. D. 2016. Data-driven confidence interval estimation incorporating prior information with an adjustment for skewed data, *The American Statistician,* 70 (3), 243-249.

Yang, Y., and He, X. 2012. Bayesian empirical likelihood for quantile regression, *The Annals of Statistics,* 40, 1102–1131.

Yu, J., Vexler, A., and Tian, L. 2010. Analyzing incomplete data subject to a threshold using empirical likelihood methods: an application to a pneumonia risk study in an ICU setting, *Biometrics,* 66, 123–130.

Zhong, X. L., and Ghosh, M. 2016. Higher-order properties of Bayesian empirical likelihood, *Electronic Journal of Statistics,* 10, 3011-3044.

# Supplementary Material for

# "The Empirical Likelihood Prior Applied To Bias Reduction of General Estimating Equations"

Albert Vexler[1], Li Zou

*Department of Biostatistics, State University of New York at Buffalo, Buffalo,*

*New York 14214, U.S.A.*

avexler@buffalo.edu    lizou@buffalo.edu

and

Alan D. Hutson

*Department of Biostatistics and Bioinformatics, Roswell Park Cancer Institute, Buffalo,*

*New York 14263, U.S.A.*

alan.hutson@roswellpark.org

**Remark A1**.

In the parametric Bayesian setting, Clarke and Yuan (2010) introduced the mutual information

$$I(\Theta; X^n) = \int\int \pi(\theta) L(X|\theta) \log\{\pi(\theta|X)/\pi(\theta)\} \mu(dX)\mu(d\theta),$$

where $\pi(\theta|X) = L(X|\theta)\pi(\theta)/\int L(X|\xi)\pi(\xi)d\xi$, $\pi(\theta)$ is the prior density function, and $L(X|\theta)$ defines the likelihood $\prod_{i=1}^{n} f(X_i|\theta)$. The notation $\mu$ generically denotes a dominated measure. To be consistent with the definitions discussed in Hartigan (1998), one can assume

---

[1] Drs. Vexler's effort was supported by the National Institutes of Health (NIH) grant 1G13LM012241-01.

$\mu(dX) = dX = \prod_{i=1}^{n} dX_i$. Clarke and Yuan (2010) examined a nonparametric version of $I(\Theta; X^n)$ of the form

$$\int\int \pi(\theta)\exp\{l(\theta)\}\log\left\{\exp\{l(\theta)\}\Big/\int \pi(\xi)\exp\{l(\xi)\}d\xi\right\}dXd\theta,$$

where the log empirical likelihood $l(\theta)$ defined in (4) heuristically plays the role of the parametric log likelihood. We refer the reader to Theorem 2.4. and p. 64 (the fourth line from the bottom) of Clarke and Yuan (2010) to a clear definition of the evaluated quantity. Unfortunately, the empirical likelihood (EL) $\exp\{l(\theta)\}$ is not a joint density function. It is well known that

$-2\{l(\theta) + n\log(n)\} \sim \left\{\sum(X_i - \theta)\right\}^2 / \sum\left(X_i - n^{-1}\sum_{i=1}^{n} X_i\right)^2$ when $G(X_i, \theta) = X_i - \theta$ in (4). Given these considerations it follows that the boundedness of

$\int \exp\{l(\theta)\}\log\left\{\exp\{l(\theta)\}\Big/\int \pi(\xi)\exp\{l(\xi)\}d\xi\right\}dX$ is a concern in general.

Unlike Clarke and Yuan (2010), in this paper we present the necessary conditions and rigorous proofs for deriving the EL prior. The following critical points can be directly associated with the results shown in Clarke and Yuan (2010).

1. In general EL functions cannot be defined for all possible values of their parameters according to the EL methodology (Owen 2001. This issue is associated with abilities to numerically evaluate the probability weights involved in many EL forms, depending on values of their parameters. For example, in a simple but common case, the log EL function for the mean has the form

$$l(\theta) = \max_{0 < p_1, \ldots, p_n < 1}\left\{\sum_{i=1}^{n}\log p_i : \sum_{i=1}^{n} p_i = 1, \sum_{i=1}^{n} X_i p_i = \theta\right\}.$$

Then the probability weights $p_i$'s, $i=1,\ldots,n$, cannot be derived when $\theta \notin \left\{\min_{i=1,\ldots,n}(X_i), \max_{i=1,\ldots,n}(X_i)\right\}$. Clarke and Yuan (2010) considered the integration of the EL function, $\exp\{l(\theta)\}$, over $\theta \in (-\infty, \infty)$. This problem cannot be corrected by taking into account the appropriate bounds of the parameter $\theta$, since in the definition of the classical EL these bounds depend on random data, $\theta \in \left\{ \min_{i=1,...,n}(X_i), \max_{i=1,...,n}(X_i) \right\}$, whereas $E[\exp\{l(\theta)\}]$ should be integrated over $\theta$ in the functional $\int \pi(\theta) E[\log\{\pi(\theta \mid X)/\pi(\theta)\}] d\theta$ and this integration cannot depend on data.

2. In several proof schemes applied in Clarke and Yuan (2010), one can detect heuristic and non-rigorous techniques that substitute in place of the necessary mathematical arguments. For example, consider the term $\Lambda_n^{-1}(\theta_n)$ analyzed in equation (2.3.13, p. 64) of Yuan and Clarke (2010). There $\Lambda_n^{-1}(\theta_n) = n^{-1} \sum_{i=1}^{n} d^2 \log p_i(\theta)/d\theta^2 \mid_{\theta=\theta_n}$, where $p_i(\theta)$, $i=1,\ldots,n$, are related to the EL function, $EL = \prod_{i=1}^{n} p_i(\theta)$ is constrained by $\sum_{i=1}^{n} p_i(\theta) = 1$, $\sum_{i=1}^{n} \{p_i(\theta) G(X_i, \theta)\} = 0$ and $\theta_n = \theta + \varpi(\hat{\theta}_n - \theta)$ with $\varpi \in (0,1)$. Thus the second order Taylor expansion of the EL function provides the relationship $\prod_{i=1}^{n} p_i(\theta) = \exp\left[\log\left\{\prod_{i=1}^{n} p_i(\hat{\theta}_n)\right\} + 0.5n(\hat{\theta}_n - \theta)^2 \Lambda_n^{-1}(\theta_n)\right]$. It is clear that $\Lambda_n^{-1}(\theta_n)$ depends on values of $\theta$ that are involved in the integral considered in Equation (2.3.13) and thus cannot be exchanged outside the integral, i.e., $\int \exp\left[-0.5n(\hat{\theta}_n - \alpha)^2 \Lambda_n^{-1}\{\alpha + \varpi(\hat{\theta}_n - \alpha)\}\right] d\alpha \neq \exp\{\Lambda_n^{-1}(\theta_n)\} \int \exp\left\{-0.5n(\hat{\theta}_n - \alpha)^2\right\} d\alpha$. Then one can conclude that result (2.3.13) provided in Clarke and Yuan (2010) is not rigorous. To obtain the asymptotic result regarding the Shannon mutual information, the second derivatives $l_i^{(2)}(\theta) = d^2 \log p_i(\theta)/d\theta^2$, $i=1,\ldots,n$, should be proven to be bounded with respect to parameter $\theta$ such that $\sum_{i=1}^{n} l_i^{(2)}(\theta) = O_p(n)$ (Vexler, Ge, and Hutson 2014a). This is assumed to be held without any justification in Clarke and Yuan (2010).

3. Clarke and Yuan (2010, pp. 64, 3rd line from the bottom) calculated the integral $\int 0.5n(\hat{\theta}_n - \theta)^2 \Lambda(\theta) \prod_{i=1}^{n} p_i(\theta) dX_1...dX_n$ with $\Lambda(\theta) = [E\{G'(X,\theta)\}]^2 [E\{G(X,\theta)G'(X,\theta)\}]^{-1}$, where $G'(X,\theta) = \partial G(X,\theta)/\partial \theta$ and $\hat{\theta}_n = \arg\max_\theta \log\left\{\prod_{i=1}^{n} p_i(\theta)\right\}$. This integral was presented as a mathematical expectation and computed with respect to the "density" function $p(X \mid \theta) = \prod_{i=1}^{n} p_i(\theta)$, see also Eq. (2.3.13) in Clarke and Yuan (2010). Since, in general, the EL function $\prod_{i=1}^{n} p_i(\theta)$ is not a proper joint density function, $\int \prod_{i=1}^{n} p_i(\theta) dX_1 ... dX_n \neq 1$ and $\int \prod_{i=1}^{n} p_i(\theta) dX_1 ... dX_n$ can be unbounded. Thus the approach of Clarke and Yuan to calculate the integral $\int 0.5n(\hat{\theta}_n - \theta)^2 \Lambda(\theta) \prod_{i=1}^{n} p_i(\theta) dX_1 ... dX_n$ is non-rigorous.

**Remark A2**.

Let the log EL ratio be $lr(\theta) = \log[\exp\{l(\theta)\} n^n]$, where

$$l(\theta) = \max_{0 < p_1, ..., p_n < 1} \left\{ \sum_{i=1}^{n} \log p_i : \sum_{i=1}^{n} p_i = 1, \sum_{i=1}^{n} (X_i p_i) = \theta \right\}.$$

Since the log EL ratio for the mean can be associated with the t-statistic (Owen 1990; Vexler et al. 2009), we begin with a consideration of the expectation of the t-statistic type object in the form $E\left[ \left\{ \sum_{i=1}^{n} (X_i - \theta) \right\}^2 / \sum_{i=1}^{n} (X_i - \theta)^2 \right]$. The statistic $\left\{ \sum_{i=1}^{n} (X_i - \theta) \right\}^2 / \sum_{i=1}^{n} (X_i - \theta)^2$ has asymptotically a $\chi_1^2$ distribution as $n \to \infty$ and $EX_1 = \theta$. Using a common technique, for all $0.5 < \varepsilon < 1$, we have

$$E\left[ \left\{ \sum_{i=1}^{n} (X_i - \theta) \right\}^2 / \sum_{i=1}^{n} (X_i - \theta)^2 \right]$$

$$= E\left[ \left\{ \sum_{i=1}^{n} (X_i - \theta) \right\}^2 \left\{ \sum_{i=1}^{n} (X_i - \theta)^2 \right\}^{-1} I\left\{ \sum_{i=1}^{n} (X_i - \theta)^2 \geq \sigma^2 n - n^\varepsilon \right\} \right]$$

$$+ E\left[ \left\{ \sum_{i=1}^{n} (X_i - \theta) \right\}^2 \left\{ \sum_{i=1}^{n} (X_i - \theta)^2 \right\}^{-1} I\left\{ \sum_{i=1}^{n} (X_i - \theta)^2 < \sigma^2 n - n^\varepsilon \right\} \right]$$

$$\leq E\left[ \left\{ \sum_{i=1}^{n} (X_i - \theta) \right\}^2 / (\sigma^2 n - n^\varepsilon) \right]$$

$$+ E\left[ n \sum_{i=1}^{n} (X_i - \theta)^2 \left\{ \sum_{i=1}^{n} (X_i - \theta)^2 \right\}^{-1} I\left\{ \sum_{i=1}^{n} (X_i - \theta)^2 \leq \sigma^2 n - n^\varepsilon \right\} \right]$$

$$= (\sigma^2 n)/(\sigma^2 n - n^\varepsilon) + n \Pr\left[\left|\sum_{i=1}^{n}\{\sigma^2 - (X_i - \theta)^2\}\right|^r > n^{\varepsilon r}\right]$$

$$\leq (\sigma^2 n)/(\sigma^2 n - n^\varepsilon) + O(n^{1+(0.5-\varepsilon)r}),$$

$$E\left[\left\{\sum_{i=1}^{n}(X_i - \theta)\right\}^2 / \sum_{i=1}^{n}(X_i - \theta)^2\right]$$

$$\geq E\left[\left\{\sum_{i=1}^{n}(X_i - \theta)\right\}^2 \left\{\sum_{i=1}^{n}(X_i - \theta)^2\right\}^{-1} I\left\{\sum_{i=1}^{n}(X_i - \theta)^2 < \sigma^2 n + n^\varepsilon\right\}\right]$$

$$\geq E\left[\left\{\sum_{i=1}^{n}(X_i - \theta)\right\}^2 / (\sigma^2 n + n^\varepsilon) I\left\{\sum_{i=1}^{n}(X_i - \theta)^2 < \sigma^2 n + n^\varepsilon\right\}\right]$$

$$= E\left\{\sum_{i=1}^{n}(X_i - \theta)\right\}^2 / (\sigma^2 n + n^\varepsilon)$$

$$- E\left[\left\{\sum_{i=1}^{n}(X_i - \theta)\right\}^2 \{\sigma^2 n + n^\varepsilon\}^{-1} I\left\{\sum_{i=1}^{n}(X_i - \theta)^2 \geq \sigma^2 n + n^\varepsilon\right\}\right]$$

$$= (\sigma^2 n)/(\sigma^2 n - n^\varepsilon) + O(n^{1+(0.5-\varepsilon)r}), \; \sigma^2 = E(X_1 - \theta)^2 \text{ and } r > (\varepsilon - 0.5)^{-1},$$

Thus we show that $E\left[\left\{\sum_{i=1}^{n}(X_i - \theta)\right\}^2 / \sum_{i=1}^{n}(X_i - \theta)^2\right] \to E\chi_1^2 = 1$, as $n \to \infty$. The classical Cauchy-Schwarz inequality $\left\{\sum_{i=1}^{n}(X_i - \theta)\right\}^2 \leq n\sum_{i=1}^{n}(X_i - \theta)^2$ plays an important role in the proof scheme above. This inequality ensures that $E\left[\left\{\sum_{i=1}^{n}(X_i - \theta)\right\}^2 / \sum_{i=1}^{n}(X_i - \theta)^2\right] < \infty$, for all $\theta$.

Regarding the expectation of the log EL ratio, one can show that $E\{lr(\theta)\} = E\left\{\sum_{i=1}^{n}\log(1 + \lambda(X_i - \theta)/n)\right\}$ where $\lambda$ is a root of $n^{-1}\sum_{i=1}^{n}[X_i/\{n + \lambda(X_i - \theta)\}] = 0$ (Owen 2001). In this case, we cannot use the Cauchy-Schwarz inequality to show $E\{lr(\theta)\}$ is bounded. To outline this concern, we use the Taylor expansion, in which $\lambda(X_i - \theta)/n$ can be expanded around 0, to note that

$$E\{lr(\theta)\} = E\left|\sum_{i=1}^{n}\log\{1 + \lambda(X_i - \theta)/n\}\right| \approx E\left\{\sum_{i=1}^{n}\lambda(X_i - \theta)/n\right\}, \text{ as } n \to \infty.$$

The appendix of our paper presents Lemma A3 that demonstrates

$$E\left\{\sum_{i=1}^{n}\lambda(X_i-\theta)/n\right\}= E\left[\left\{\sum_{i=1}^{n}(X_i-\theta)\right\}^2 \Big/\left\{n\sum_{i=1}^{n}(X_i-\theta)^2 p_i\right\}\right],$$

where it is incorrect in general to state that $\left\{\sum_{i=1}^{n}(X_i-\theta)\right\}^2 \leq n^2\sum_{i=1}^{n}(X_i-\theta)^2 p_i$ for certain values of $\theta$, especially if values of $\theta$ are close to $\min_{i=1,...,n}(X_i)$ or $\max_{i=1,...,n}(X_i)$. For example, when $\theta$ is very close to $\min_{i=1,...,n}(X_i)$, then

$$E\left[\left\{\sum_{i=1}^{n}(X_i-\theta)\right\}^2 \Big/\left\{n\sum_{i=1}^{n}(X_i-\theta)^2 p_i\right\}\right] \approx E\left[\left\{\sum_{i=1}^{n}(X_i-\theta)\right\}^2 \Big/\left\{n\left(\min_{i=1,...,n}(X_i)-\theta\right)^2\right\}\right]$$

may be unbounded. In order to implement a similar role as that of the Cauchy-Schwarz inequality in the t-statistic context shown above, we redefine the log EL ratio in this paper as

$$lr_e(\theta) = lr(\theta)I\{B_n(\theta)\} + \log(D)I\{B_n^c(\theta)\}, \ B_n(\theta) = \{w_1 > M, w_2 > M\},$$

where $w_1 = \sum_{i=1}^{n}\{G(X_i,\theta)\}^2 I\{G(X_i,\theta)<0\}$, $w_2 = \sum_{i=1}^{n}\{G(X_i,\theta)\}^2 I\{G(X_i,\theta)>0\}$, and $M$ is a fixed constant. It is clear that $E\{lr_e(\theta)\}$ is bounded for all $\theta$; for details see the proofs of Lemmas A1 and A5 in the Supplementary Material.

**Proofs**

*Proof of Proposition 1.*

*Lemma A1.* Assume that, for $\gamma > 0$, $E|G(X_1,\theta)|^{4+\gamma} < \infty$. Then, defining $b_n = n^{1-\varepsilon}$ with $0 < \varepsilon < 1$, we have

$$\int_0^{b_n}\Pr\{-2lr(\theta) > x, B_n(\theta)\}dx = 1, \text{ as } n \to \infty.$$

*Proof of Lemma A1.*

Note that $b_n \Pr(B_n^c) = o(1)$ as $n \to \infty$, since the Chebyshev inequality provides that

$$\Pr\left[\sum_{i=1}^{n}G(X_i,\theta)^2 I\{G(X_i,\theta)<0\} < M\right] \leq E\left|\sum_{i=1}^{n}[q - G(X_i,\theta)^2 I\{G(X_i,\theta)<0\}]\right|^2 (nq-M)^{-2}$$

$$= O(n^{-1.5}),$$

where $q = E[\{G(X_1,\theta)\}^2 I\{G(X_1,\theta) < 0\}]$.

By a result in DiCiccio, Hall, and Romano (1991, p. 1055), we have

$$\int_0^{b_n} \Pr\{-2lr(X \mid \theta) > x\}dx = \int_0^{b_n} [\Pr\{\chi_1^2 > x\} + O(n^{-1})]dx = E\chi_1^2 + o(1) = 1 + o(1), \quad n \to \infty.$$

It follows that

$$\int_0^{b_n} \Pr\{-2lr(X \mid \theta) > x, B_n(\theta)\}dx \le \int_0^{b_n} \Pr\{-2lr(X \mid \theta) > x\}dx = 1 + o(1),$$

and

$$\int_0^{b_n} \Pr\{-2lr(\theta) > x, B_n(\theta)\}dx \ge \int_0^{b_n} \Pr\{-2lr(\theta) > x\}dx - \int_0^{b_n} \Pr\{B_n^c(\theta)\}dx$$

$$= \int_0^{b_n} \Pr\{-2lr(\theta) > x\}dx - b_n P\{B_n^c(\theta)\} = 1 + o(1).$$

This completes the proof of Lemma A1.

*Lemma A2.* We have that $\lambda \ge 0$ if and only if $\sum_{i=1}^{n} G(X_i,\theta) \ge 0$, and $\lambda < 0$ if and only if

$$\sum_{i=1}^{n} G(X_i,\theta) < 0.$$

*Proof of Lemma A2.*

The forms, $p_i = \{n + \lambda G(X_i,\theta)\}^{-1}$, $i = 1,..,n$, and $n^{-n} \ge \prod_{i=1}^{n} p_i$ imply that

$$0 \le -lr(\theta) = \log\left(n^{-n} / \prod_{i=1}^{n} p_i\right) = \sum_{i=1}^{n} \log\{1 + \lambda G(X_i,\theta)/n\}.$$

Using the inequality $\log(1+s) \le s$ for $s > -1$, we obtain

$$-lr(\theta) = \sum_{i=1}^{n} \log\{1 + \lambda G(X_i,\theta)/n\}$$

$$\leq \sum_{i=1}^{n} \lambda G(X_i,\theta)/n = \lambda \sum_{i=1}^{n} G(X_i,\theta)/n.$$

This completes the proof of Lemma A2.

*Lemma A3.* The Lagrange multiplier $\lambda$ satisfies $\lambda = \sum_{i=1}^{n} G(X_i,\theta) / \sum_{i=1}^{n} \{G(X_i,\theta)\}^2 p_i$.

*Proof of Lemma A3.*

The constraint $\sum_{i=1}^{n} G(X_i,\theta) p_i = 0$ with $p_i = \{n + \lambda G(X_i,\theta)\}^{-1}$, $i=1,...,n$, in (4) implies that

$$\sum_{i=1}^{n} G(X_i,\theta) = \sum_{i=1}^{n} \{G(X_i,\theta)(1-p_i)\} + \sum_{i=1}^{n} \{G(X_i,\theta) p_i\}$$

$$= \sum_{i=1}^{n} G(X_i,\theta)(1-p_i) = \sum_{i=1}^{n} G(X_i,\theta) \left\{ \frac{n + \lambda G(X_i,\theta) - 1}{n + \lambda G(X_i,\theta)} \right\} = \lambda \sum_{i=1}^{n} \{G(X_i,\theta)\}^2 p_i.$$

This completes the proof of Lemma A3.

*Lemma A4.* If $\lambda \geq 0$, we have $0 \leq \lambda \leq n \sum_{i=1}^{n} G(X_i,\theta) \left[ \sum_{i=1}^{n} \{G(X_i,\theta)\}^2 I\{G(X_i,\theta) < 0\} \right]^{-1}$; if $\lambda < 0$,

we have $n \sum_{i=1}^{n} G(X_i,\theta) \left[ \sum_{i=1}^{n} \{G(X_i,\theta)\}^2 I\{G(X_i,\theta) > 0\} \right]^{-1} \leq \lambda < 0$.

*Proof of Lemma A4.*

Having $\lambda \geq 0$, we obtain

$$\sum_{i=1}^{n} \{G(X_i,\theta)\}^2 p_i \geq \sum_{i=1}^{n} \left[ G(X_i,\theta)^2 p_i I\{G(X_i,\theta) < 0\} \right]$$

$$= \sum_{i=1}^{n} \left[ \{G(X_i,\theta)\}^2 \frac{I\{G(X_i,\theta) < 0\}}{n + \lambda G(X_i,\theta)} \right] \geq \sum_{i=1}^{n} \left[ \{G(X_i,\theta)\}^2 \frac{1}{n} I\{G(X_i,\theta) < 0\} \right],$$

where $p_i = \{n + \lambda G(X_i,\theta)\}^{-1}$, $i=1,...,n$.

Applying this result and Lemma A2 to Lemma A3 yields

$$0 \leq \lambda < n \sum_{i=1}^{n} G(X_i,\theta) \left[ \sum_{i=1}^{n} \{G(X_i,\theta)\}^2 I(G(X_i,\theta) < 0) \right]^{-1}.$$

It follows similarly that when $\lambda < 0$,

$$n\sum_{i=1}^{n}G(X_i,\theta)\left[\sum_{i=1}^{n}\{G(X_i,\theta)\}^2 I\{G(X_i,\theta)>0\}\right]^{-1} < \lambda < 0.$$

This completes the proof of Lemma A4.

*Lemma A5.* Assume that for $\gamma > 0$, $E|G(X_1,\theta)|^{8+\gamma} < \infty$. Then

$$\int_{b_n}^{\infty} \Pr\{-2lr(X\mid\theta) > x, B_n(\theta)\}dx = o(1) \text{ as } n \to \infty.$$

*Proof of Lemma A5.*

Since $\log(1+s) \le s$ for $s > -1$, we apply Lemmas A2 and A4 to obtain the following inequality

$$\int_{b_n}^{\infty}\Pr\{-2lr(\theta) > x, B_n(\theta)\}dx = \int_{b_n}^{\infty}\Pr\left\{\sum_{i=1}^{n}\log\{1+\lambda G(X_i,\theta)n^{-1}\} > x/2, B_n(\theta)\right\}dx$$

$$\le \int_{b_n}^{\infty}\Pr\left\{n^{-1}\sum_{i=1}^{n}\lambda G(X_i,\theta) > x/2, B_n(\theta)\right\}dx$$

$$= \int_{b_n}^{\infty}\Pr\left\{\lambda I(\lambda\ge 0)n^{-1}\sum_{i=1}^{n}G(X_i,\theta)+\lambda I(\lambda<0)n^{-1}\sum_{i=1}^{n}G(X_i,\theta)I(\lambda<0) > \frac{x}{2}, B_n(\theta)\right\}dx$$

$$\le \int_{b_n}^{\infty}\Pr\left\{\left(\sum_{i=1}^{n}G(X_i,\theta)\right)^2\left(\sum_{i=1}^{n}G(X_i,\theta)^2 I(G(X_i,\theta)<0)\right)^{-1} > x/4, B_n(\theta)\right\}dx$$

$$+ \int_{b_n}^{\infty}\Pr\left\{\left(\sum_{i=1}^{n}G(X_i,\theta)\right)^2\left(\sum_{i=1}^{n}G(X_i,\theta)^2 I(G(X_i,\theta)>0)\right)^{-1} > x/4, B_n(\theta)\right\}dx.$$

Next we show that

$$\int_{b_n}^{\infty}\Pr\left[\left\{\sum_{i=1}^{n}G(X_i,\theta)\right\}^2\left[\sum_{i=1}^{n}G(X_i,\theta)^2 I\{G(X_i,\theta)<0\}\right]^{-1} > x/4, B_n(\theta)\right]dx = o(1),$$

$$\int_{b_n}^{\infty}\Pr\left[\left\{\sum_{i=1}^{n}G(X_i,\theta)\right\}^2\left[\sum_{i=1}^{n}G(X_i,\theta)^2 I\{G(X_i,\theta)>0\}\right]^{-1} > x/4, B_n(\theta)\right]dx = o(1).$$

Define $q = E[\{G(X_1,\theta)\}^2 I\{G(X_1,\theta)<0\}]$, $\varepsilon \in (0,q)$ and the event

$$W_n(\theta) = \left\{\sum_{i=1}^{n}U(X_i,\theta) < qn - \varepsilon n\right\}, \text{ where } U(X_i,\theta) = \{G(X_i,\theta)\}^2 I\{G(X_i,\theta)<0\}.$$

Then

$$\int_{b_n}^{\infty} \Pr\left\{\left(\sum_{i=1}^{n} G(X_i,\theta)\right)^2 \left(\sum_{i=1}^{n} \{G(X_i,\theta)\}^2 I(G(X_i,\theta)<0)\right)^{-1} > \frac{x}{4}, B_n(\theta)\right\} dx$$

$$\leq \int_{b_n}^{\infty} \Pr\left\{\left(\sum_{i=1}^{n} G(X_i,\theta)\right)^2 \left(\sum_{i=1}^{n} U(X_i,\theta)\right)^{-1} > \frac{x}{4}, \sum_{i=1}^{n} U(X_i,\theta) > M\right\} dx$$

$$= \int_{b_n}^{\infty} \Pr\left\{\left(\sum_{i=1}^{n} G(X_i,\theta)\right)^2 \left(\sum_{i=1}^{n} U(X_i,\theta)\right)^{-1} > \frac{x}{4}, \sum_{i=1}^{n} U(X_i,\theta) > M, W_n(\theta)\right\} dx$$

$$+ \int_{b_n}^{\infty} \Pr\left\{\left(\sum_{i=1}^{n} G(X_i,\theta)\right)^2 \left(\sum_{i=1}^{n} U(X_i,\theta)\right)^{-1} > \frac{x}{4}, \sum_{i=1}^{n} U(X_i,\theta) > M, W_n^C(\theta)\right\} dx$$

$$\leq \int_{b_n}^{\infty} \Pr\left\{\left(\sum_{i=1}^{n} G(X_i,\theta)\right)^2 M^{-1} > x, \sum_{i=1}^{n} U(X_i,\theta) < qn - \varepsilon n\right\} dx$$

$$+ \int_{b_n}^{\infty} \Pr\left\{\left(\sum_{i=1}^{n} G(X_i,\theta)\right)^2 (qn - \varepsilon n)^{-1} > x\right\} dx.$$

Let $\gamma > \varepsilon$, it is clear that Chebyshev inequality implies $\Pr\left\{\sum_{i=1}^{n} U(X_i,\theta) < qn - \varepsilon n\right\} = O(n^{-2-\gamma})$, provided that $E|G(X_1,\theta)|^{8+\gamma} < \infty$ as $n \to \infty$.

Then

$$\int_{b_n}^{\infty} \Pr\left\{\left(\sum_{i=1}^{n} G(X_i,\theta)\right)^2 M^{-1} > x, \sum_{i=1}^{n} U(X_i,\theta) < qn - \varepsilon n\right\} dx$$

$$\leq \int_{b_n}^{\infty} \left[\Pr\left(\sum_{i=1}^{n} U(X_i,\theta) < qn - \varepsilon n\right) \Pr\left\{\left(\sum_{i=1}^{n} G(X_i,\theta)\right)^2 M^{-1} > x\right\}\right]^{1/2} dx$$

$$\leq \int_{b_n}^{\infty} \left[O(n^{-2-\gamma}) E\left(\sum_{i=1}^{n} G(X_i,\theta)\right)^6 \frac{1}{M^3 x^3}\right]^{1/2} dx$$

$$= O(n^{-1-0.5\gamma}) \int_{b_n}^{\infty} \{n^3 M^{-3} x^{-3}\}^{1/2} dx = O(n^{-0.5\gamma + 0.5\varepsilon}).$$

Thus we obtain

$$\int_{b_n}^{\infty} \Pr\left[\left\{\sum_{i=1}^{n} G(X_i,\theta)\right\}^2 \left[\sum_{i=1}^{n} G(X_i,\theta)^2 I\{G(X_i,\theta)<0\}\right]^{-1} > x/4, B_n(\theta)\right] dx = o(1).$$

In a similar manner to the considerations above, one can show that

$$\int_{b_n}^{\infty} \Pr\left[\left\{\sum_{i=1}^{n} G(X_i,\theta)\right\}^2 \left[\sum_{i=1}^{n} G(X_i,\theta)^2 I\{G(X_i,\theta)>0\}\right]^{-1} > x/4, B_n(\theta)\right] dx = o(1).$$

The proof of Lemma A5 is complete.

*Lemma A6.* Assume that $|\partial G(x,\theta)/\partial \theta|$ is bounded by some function $Q(x)$ for $\theta \in [\hat{\theta} - \varphi_n n^{-1/2}, \hat{\theta} + \varphi_n n^{-1/2}]$ where $\varphi_n = n^{1/6-\beta}$ with $\beta \in (0,1/6)$. Then

$$0 < -\log\left\{\int_{\hat{\theta}-\varphi_n n^{-1/2}}^{\hat{\theta}+\varphi_n n^{-1/2}} e^{lr_e(\zeta)} \pi(\zeta) d\zeta\right\} \leq \left\{2\varphi_n^2 n^{-1}\left(\sum_{i=1}^{n} Q(X_i)\right)^2 / M - \log(D_n)\right\} - \log\left\{\int_{\hat{\theta}-\varphi_n n^{-1/2}}^{\hat{\theta}+\varphi_n n^{-1/2}} \pi(\zeta) d\zeta\right\},$$

where $D_n = \exp(-c_0 n)$ and $c_0$ is a positive constant.

*Proof of Lemma A6.*

In the proof of Lemma A4, we demonstrated that

$$p_i = \{n + \lambda G(X_i,\theta)\}^{-1} \geq n^{-1} I\{G(X_i,\theta)<0\} \text{ for } \lambda \geq 0, \text{ and}$$

$$p_i = \{n + \lambda G(X_i,\theta)\}^{-1} \geq n^{-1} I\{G(X_i,\theta)>0\} \text{ for } \lambda < 0, \text{ for } i=1,\ldots,n.$$

Then taking into account the definitions of $lr_e(\theta)$ and $B_n(\theta)$, and Lemma A3, we obtain

$$lr(\theta)I\{B_n(\theta)\} = lr(\theta)I\{B_n(\theta)\}I(\lambda \geq 0) + lr(\theta)I\{B_n(\theta)\}I(\lambda < 0)$$

$$\geq -\frac{\left\{\sum_{i=1}^{n} G(X_i,\theta)\right\}^2}{n \sum_{i=1}^{n} \{G(X_i,\theta)\}^2 p_i} I\{B_n(\theta)\}I(\lambda \geq 0) - \frac{\left\{\sum_{i=1}^{n} G(X_i,\theta)\right\}^2}{n \sum_{i=1}^{n} \{G(X_i,\theta)\}^2 p_i} I\{B_n(\theta)\}I(\lambda < 0)$$

$$\geq -\frac{\left\{\sum_{i=1}^{n} G(X_i,\theta)\right\}^2 I\{B_n(\theta)\}I(\lambda \geq 0)}{\sum_{i=1}^{n} \{G(X_i,\theta)\}^2 I\{G(X_i,\theta)<0\}} - \frac{\left\{\sum_{i=1}^{n} G(X_i,\theta)\right\}^2 I\{B_n(\theta)\}I(\lambda < 0)}{\sum_{i=1}^{n} \{G(X_i,\theta)\}^2 I\{G(X_i,\theta)>0\}}$$

$$\geq -\frac{2\left\{\sum_{i=1}^{n} G(X_i, \theta)\right\}^2}{M}.$$

The Taylor expansion of $\sum_{i=1}^{n} G(X_i, \zeta)$ with $\zeta$ around $\hat{\theta}$, when $\zeta \in \left[\hat{\theta} - \varphi_n n^{-1/2}, \hat{\theta} + \varphi_n n^{-1/2}\right]$, provides

$$\sum_{i=1}^{n} G(X_i, \zeta) = \sum_{i=1}^{n} G(X_i, \hat{\theta}) + (\zeta - \hat{\theta}) \sum_{i=1}^{n} G'(X_i, \bar{\theta}_i),$$

where $G'(X_i, \zeta) = \partial G(X_i, \zeta) / \partial \zeta$, $\bar{\theta}_i = \hat{\theta} + \omega_i (\zeta - \hat{\theta})$ and $\omega_i \in (0,1)$ for $i=1,\ldots,n$.

Note that, by the definition of $\hat{\theta}$, we have $\sum_{i=1}^{n} G(X_i, \hat{\theta}) = 0$. Assume that $|\partial G(x, \theta) / \partial \theta|$ is bounded by some function $Q(x)$ for $\theta \in \left[\hat{\theta} - \varphi_n n^{-1/2}, \hat{\theta} + \varphi_n n^{-1/2}\right]$, we have

$$\left|\sum_{i=1}^{n} G(X_i, \zeta)\right| \leq \varphi_n n^{-1/2} \sum_{i=1}^{n} Q(X_i),$$

for $\zeta \in \left[\hat{\theta} - \varphi_n n^{-1/2}, \hat{\theta} + \varphi_n n^{-1/2}\right]$.

Then for $\zeta \in \left[\hat{\theta} - \varphi_n n^{-1/2}, \hat{\theta} + \varphi_n n^{-1/2}\right]$, we obtain

$$lr(\zeta) I\{B_n(\zeta)\} + \log(D_n) I\{B_n^c(\zeta)\} \geq -2\varphi_n^2 n^{-1} \left(\sum_{i=1}^{n} Q(X_i)\right)^2 / M + \log(D_n).$$

Thus by (6), one can easily obtain that

$$0 < -\log\left\{\int_{\hat{\theta}-\varphi_n n^{-1/2}}^{\hat{\theta}+\varphi_n n^{-1/2}} e^{lr_e(\theta)} \pi(\zeta) d\zeta\right\} \leq \left\{2\varphi_n^2 n^{-1} \left(\sum_{i=1}^{n} Q(X_i)\right)^2 / M - \log(D_n)\right\} - \log\left\{\int_{\hat{\theta}-\varphi_n n^{-1/2}}^{\hat{\theta}+\varphi_n n^{-1/2}} \pi(\zeta) d\zeta\right\}.$$

This completes the proof of Lemma A6.

*Lemma A7.* Assume that $|\partial G(x, \theta) / \partial \theta|$ is bounded by some function $Q(x)$ with $E\{Q(X_1)^4\} < \infty$ for $\theta \in \left[\hat{\theta} - \varphi_n n^{-1/2}, \hat{\theta} + \varphi_n n^{-1/2}\right]$ where $\varphi_n = n^{1/6-\beta}$ with $\beta \in (0, 1/6)$. Then

$$-E\left[\log\left\{\int_{\hat{\theta}-\varphi_n n^{-1/2}}^{\hat{\theta}+\varphi_n n^{-1/2}} \exp\{lr_e(\zeta)\}\pi(\zeta)d\zeta + d_n\right\}I(R_n \geq d_n)\right] \to 0,$$

where $R_n = \int_{-\infty}^{\hat{\theta}-\varphi_n n^{-1/2}} \exp\{lr_e(\zeta)\}\pi(\zeta)d\zeta + \int_{\hat{\theta}+\varphi_n n^{-1/2}}^{\infty} \exp\{lr_e(\zeta)\}\pi(\zeta)d\zeta$ and $d_n = \exp(-n^{\xi})$ with $0 < \zeta < 1/3 - 2\beta$.

*Proof of Lemma A7.*

Since $-\log\left\{\int_{\hat{\theta}-\varphi_n n^{-1/2}}^{\hat{\theta}+\varphi_n n^{-1/2}} \exp\{lr_e(\zeta)\}\pi(\zeta)d\zeta + d_n\right\} \geq 0$, the Cauchy-Schwarz inequality implies

$$0 \leq -E\left[\log\left\{\int_{\hat{\theta}-\varphi_n n^{-1/2}}^{\hat{\theta}+\varphi_n n^{-1/2}} \exp\{lr_e(\zeta)\}\pi(\zeta)d\zeta + d_n\right\}I(R_n \geq d_n)\right]$$

$$\leq \left[E\left\{\log\left(\int_{\hat{\theta}-\varphi_n n^{-1/2}}^{\hat{\theta}+\varphi_n n^{-1/2}} \exp\{lr_e(\zeta)\}\pi(\zeta)d\zeta + d_n\right)\right\}^2\right]^{1/2}\{\Pr(R_n \geq d_n)\}^{1/2}. \quad (S.1)$$

Because $E\{Q(X_1)^4\} < \infty$, Lemma A6 leads to

$$\left[E\left\{\log\left(\int_{\hat{\theta}-\varphi_n n^{-1/2}}^{\hat{\theta}+\varphi_n n^{-1/2}} \exp\{lr_e(\zeta)\}\pi(\zeta)d\zeta + d_n\right)\right\}^2\right]^{1/2} = O(n^{4/3-2\beta}).$$

In the next stage, we will evaluate the remainder term $R_n$ in $\Pr(R_n \geq d_n)$. According to Lemma A1 in Vexler, Ge, and Hutson (2014a, p. 3 in the Supplementary Material), the function $lr(\theta)$ increases for $\theta < \hat{\theta}$ and the function $lr(\theta)$ decreases for $\theta > \hat{\theta}$. Then it follows that

$$0 \leq \int_{-\infty}^{\hat{\theta}-\varphi_n n^{-1/2}} e^{lr(\zeta)}\pi(\zeta)d\zeta \leq e^{lr(\hat{\theta}-\varphi_n n^{-1/2})} \text{ and } 0 \leq \int_{\hat{\theta}+\varphi_n n^{-1/2}}^{\infty} e^{lr(\zeta)}\pi(\zeta)d\zeta \leq e^{lr(\hat{\theta}+\varphi_n n^{-1/2})}. \quad (S.2)$$

In order to evaluate $lr(\hat{\theta} - \varphi_n n^{-1/2})$, one can use the techniques shown in Vexler, Ge, and Hutson (2014a, p. 6 of the Supplementary Material). To this end, we first derive a bound for $\lambda(\theta)$ when $\theta = \hat{\theta} - \varphi_n n^{-1/2}$. Since $\lambda(\theta)$ is a root of the equation $n^{-1}\sum_{i=1}^{n}G(X_i,\theta)/\{1+n^{-1}\lambda G(X_i,\theta)\}=0$, we define the function

$$L(\lambda) = n^{-1}\sum_{i=1}^{n}G(X_i,\hat{\theta}-\varphi_n n^{-1/2})/\{1+n^{-1}\lambda G(X_i,\hat{\theta}-\varphi_n n^{-1/2})\}.$$

Then

$$L(\lambda) = n^{-1}\sum_{i=1}^{n}\frac{G(X_i,\hat{\theta}-\varphi_n n^{-1/2})}{\{1+n^{-1}\lambda G(X_i,\hat{\theta}-\varphi_n n^{-1/2})\}}$$

$$= n^{-1}\sum_{i=1}^{n}G(X_i,\hat{\theta}-\varphi_n n^{-1/2}) - n^{-1}\sum_{i=1}^{n}\frac{n^{-1}\lambda G(X_i,\hat{\theta}-\varphi_n n^{-1/2})^2}{\{1+n^{-1}\lambda G(X_i,\hat{\theta}-\varphi_n n^{-1/2})\}}. \quad (S.3)$$

Let $\lambda_c = n^{2/3}\tau_n^{-1}$, where $\tau_n = n^{\nu}$, $0 < \nu < \beta < 1/6$, and a Taylor expansion that

$$\sum_{i=1}^{n}G(X_i,\hat{\theta}-\varphi_n n^{-1/2}) = \sum_{i=1}^{n}G(X_i,\hat{\theta}) - \varphi_n n^{-1/2}\sum_{i=1}^{n}\partial G(X_i,\hat{\theta}_{1i})/\partial\zeta, \quad (S.4)$$

where $\hat{\theta}_{1i} = \hat{\theta} + \omega_{1i}(\zeta - \hat{\theta})$ and $\omega_{1i} \in (0,1)$ for $i=1,\ldots,n$, substituting $\lambda_c = n^{2/3}\tau_n^{-1}$ and (S.4) into (S.3) yields

$$n^{1/2}L(\lambda_c) = \varphi_n\left(n^{-1}\sum_{i=1}^{n}\partial G(X_i,\hat{\theta}_{1i})/\partial\theta\right) - \frac{n^{1/6}}{\tau_n}\frac{1}{n}\sum_{i=1}^{n}\frac{G(X_i,\hat{\theta}-\varphi_n n^{-1/2})^2}{\{1+n^{-1/3}\tau_n^{-1}G(X_i,\hat{\theta}-\varphi_n n^{-1/2})\}},$$

Since $n^{-1/3}\tau_n^{-1}G(X_i,\hat{\theta}-\varphi_n n^{-1/2}) = O_p(1)$ (Owen 1990), we have

$$n^{1/2}L(\lambda_c) = \varphi_n\left(n^{-1}\sum_{i=1}^{n}\partial G(X_i,\hat{\theta}_{1i})/\partial\theta\right) - \frac{n^{1/6}}{\tau_n}\frac{1}{n}\sum_{i=1}^{n}\frac{G(X_i,\hat{\theta}-\varphi_n n^{-1/2})^2}{\{1+O_p(1)\}}.$$

Now it follows that $n^{1/2}L(\lambda_c) \to -\infty$, as $n \to \infty$. In a similar manner $n^{1/2}L(-\lambda_c) \to +\infty$, as $n \to \infty$. Thus, the solution, $\lambda_0$, of equation $n^{1/2}L(\lambda_0)=0$ belongs to the interval $(-\lambda_c,\lambda_c)$, i.e.

$\lambda_0 = O_p(n^{2/3} \tau_n^{-1})$. Note that this bound for $\lambda_0$ can also be obtained via using (S.4) and the exact bounds for $\lambda$ shown in Lemma A4. This result will be used to derive an expression of $\lambda_0$.

Following the same technique of a Taylor expansion of $L(\lambda_0) = 0$ that is shown in Vexler, Ge, and Hutson (2014a, pp. 6-7 of the Supplementary Material) and (S.4), we then obtain

$$\sum_{i=1}^{n} G(X_i, \hat{\theta} - \varphi_n n^{-1/2}) \left\{ 1 - n^{-1} \lambda_0 G(X_i, \hat{\theta} - \varphi_n n^{-1/2}) + \frac{n^{-2} \lambda_0 G(X_i, \hat{\theta} - \varphi_n n^{-1/2})^2}{(1 + \theta_{0i})^2} \right\} = 0, \quad (S.5)$$

where $\hat{\theta}_{0i} = \hat{\theta} - \omega_{0i} \varphi_n n^{-1/2}$ and $\omega_{0i} \in (0,1)$ for $i=1,...,n$. Since $\lambda_0 = O_p(n^{2/3} \tau_n^{-1})$ and (S.4), it follows that the approximate solution based on solving (S.5) is given by

$$\lambda_0 = \frac{-\varphi_n n^{-1/2} \sum_{i=1}^{n} G'(X_i, \hat{\theta}_{1i})}{n^{-1} \sum_{i=1}^{n} G(X_i, \hat{\theta} - \varphi_n n^{-1/2})^2} + \frac{O(n^{1/3})}{\tau_n^2}, \quad (S.6)$$

where $G'(X_i, \theta) = \partial G(X_i, \theta) / \partial \theta$, and $\hat{\theta}_{1i} = \hat{\theta} + \omega_{1i}(\zeta - \hat{\theta})$ with $\omega_{1i} \in (0,1)$ for $i=1,...,n$.

Applying the Taylor expansion, (S.4) and (S.6) to $lr(\hat{\theta} - \varphi_n n^{-1/2})$, we then have

$$lr(\hat{\theta} - \varphi_n n^{-1/2}) = -\sum_{i=1}^{n} n^{-1} \lambda_0 G(X_i, \hat{\theta} - \varphi_n n^{-1/2}) + \frac{1}{2} \sum_{i=1}^{n} n^{-1} \lambda_0 G(X_i, \hat{\theta} - \varphi_n n^{-1/2}) + O(n^{-3\gamma})$$

$$= \frac{-\varphi_n^2 \left\{ n^{-1} \sum_{i=1}^{n} G'(X_i, \hat{\theta}_{1i}) \right\}^2}{2 n^{-1} \sum_{i=1}^{n} G(X_i, \hat{\theta} - \varphi_n n^{-1/2})^2} + O_p(n^{-3\nu}). \quad (S.7)$$

Similarly to the analysis of $lr(\hat{\theta} - \varphi_n n^{-1/2})$, $lr(\hat{\theta} + \varphi_n n^{-1/2})$ has the following expression

$$lr(\hat{\theta} + \varphi_n n^{-1/2}) = \frac{-\varphi_n^2 \left\{ n^{-1} \sum_{i=1}^{n} G'(X_i, \hat{\theta}_{2i}) \right\}^2}{2 n^{-1} \sum_{i=1}^{n} G(X_i, \hat{\theta} - \varphi_n n^{-1/2})^2} + O_p(n^{-3\nu}), \quad (S.8)$$

where $\hat{\theta}_{2i} = \hat{\theta} + \omega_{2i}(\zeta - \hat{\theta})$ and $\omega_{2i} \in (0,1)$ for $i=1,\ldots,n$.

We remark that the results above are consistent with those shown in Qin and Lawless (1994, pp.316-317), where it is demonstrated that

$$\lambda_0 = O(n^{2/3}), \text{ and } lr(\hat{\theta} - \varphi_n n^{-1/2}) \leq -a_0 n^{1/3}, \text{ (a.s.)}$$

where $a_0$ is some positive constant.

Now consider the probability $\Pr(R_n \geq d_n)$

$$\Pr(R_n \geq d_n) = \Pr\left[\int_{-\infty}^{\hat{\theta}-\varphi_n n^{-1/2}} \exp\{lr(\zeta)I\{B_n(\zeta)\} + \log(D_n)I\{B_n^c(\zeta)\}\}\pi(\zeta)d\zeta\right.$$

$$\left. + \int_{\hat{\theta}+\varphi_n n^{-1/2}}^{\infty} \exp\{lr(\zeta)I\{B_n(\zeta)\} + \log(D_n)I\{B_n^c(\zeta)\}\}\pi(\zeta)d\zeta \geq d_n\right]$$

$$= \Pr\left[\int_{-\infty}^{\hat{\theta}-\varphi_n n^{-1/2}} \exp\{lr(\zeta)\}\pi(\zeta)I\{B_n(\zeta)\}d\zeta + \int_{-\infty}^{\hat{\theta}-\varphi_n n^{-1/2}} D_n\pi(\zeta)I\{B_n^c(\zeta)\}d\zeta\right.$$

$$\left. + \int_{\hat{\theta}+\varphi_n n^{-1/2}}^{\infty} \exp\{lr(\zeta)\}\pi(\zeta)I\{B_n(\zeta)\}d\zeta + \int_{\hat{\theta}+\varphi_n n^{-1/2}}^{\infty} D_n\pi(\zeta)I\{B_n^c(\zeta)\}d\zeta \geq d_n\right]$$

$$\leq \Pr\left[\exp\{lr(\hat{\theta}-\varphi_n n^{-1/2})\} + \exp\{lr(\hat{\theta}+\varphi_n n^{-1/2})\} \geq d_n - \exp(-c_0 n)\right],$$

where the inequality (S.1) is employed, $d_n = \exp(-n^\xi)$ with $0 < \xi < 1/3 - 2\beta$ and $D_n = \exp(-c_0 n)$.

Combining (S.7) and (S.8) in the inequality of $\Pr(R_n \geq d_n)$, and taking into account $\sum_{i=1}^n G'(X,\theta)$ is bounded for $\theta \in [\hat{\theta} - \varphi_n n^{-1/2}, \hat{\theta} + \varphi_n n^{-1/2}]$, we conclude that $O(n^{4/3-2\beta})\{P(R_n \geq d_n)\}^{1/2} \to 0$ as $n \to \infty$, and then using (S.1) we complete the proof of Lemma A7.

*Lemma A8.* Assume that, for some $\gamma > 0$ and all $\theta$. $E|G(X_1,\theta)|^{8+\gamma} < \infty$ and the following conditions are satisfied: $|\partial G(x,\theta)/\partial\theta|$ and $|\partial^2 G(x,\theta)/\partial\theta^2|$ are bounded by some function $Q(x)$ with $E\{Q(X_1)^4\} < \infty$, for $\theta \in [\hat{\theta} - \varphi_n n^{-1/2}, \hat{\theta} + \varphi_n n^{-1/2}]$, where $\varphi_n = n^{1/6-\beta}$, $\beta \in (0, 1/6)$. Define $\pi(\theta)$ to be a twice continuously differentiable on $\theta \in [\hat{\theta} - \varphi_n n^{-1/2}, \hat{\theta} + \varphi_n n^{-1/2}]$ prior density function. Then

$$-E\left[\log\left\{\int \pi(\zeta)\exp\{lr_e(\zeta)\}d\zeta\right\}\right] = -\log\left[\pi(\theta)\{2\pi\sigma^2(\theta)/n\}^{1/2}\right] + o(1),$$

where $\sigma^2(\theta) = E\{G(X,\theta)\}^2 / \{EG'(X,\theta)\}^2$ and $G'(X,\theta) = \partial G(X,\theta)/\partial\theta$.

*Proof of Lemma A8.*

Result (A.4) has the form

$$-E\left[\log\left\{\int_{\hat{\theta}-\varphi_n n^{-1/2}}^{\hat{\theta}+\varphi_n n^{-1/2}} \exp\{lr_e(\zeta)\}\pi(\zeta)d\zeta + d_n\right\}\right] + o(1) \leq -E\left[\log\left\{\int \exp\{lr_e(\zeta)\}\pi(\zeta)d\zeta\right\}\right]$$

$$\leq -E\left[\log\left\{\int_{\hat{\theta}-\varphi_n n^{-1/2}}^{\hat{\theta}+\varphi_n n^{-1/2}} \exp\{lr_e(\zeta)\}\pi(\zeta)d\zeta\right\}\right],\qquad (S.9)$$

where $\hat{\theta}$ satisfies $n^{-1}\sum_{i=1}^n G(X_i, \hat{\theta}) = 0$.

To prove Lemma A8, we focus on the component $E\left[\log\left\{\int_{\hat{\theta}-\varphi_n n^{-1/2}}^{\hat{\theta}+\varphi_n n^{-1/2}} \exp\{lr_e(\zeta)\}\pi(\zeta)d\zeta\right\}\right]$ at (S.9) to show the inequality

$$-E\left[\log\left\{\int_{\hat{\theta}-\varphi_n n^{-1/2}}^{\hat{\theta}+\varphi_n n^{-1/2}} \exp\{lr_e(\zeta)\}\pi(\zeta)d\zeta\right\}\right] \leq -\log\left[\pi(\theta)\{2\pi\sigma^2(\theta)/n\}^{1/2}\right] + o(1),$$

as $n \to \infty$.

By the definition of $lr_e(\zeta)$ presented in (6) where $\exp\{\log(D_n)\}I\{B_n^c(\zeta)\} \geq 0$, we have

$$-E\left[\log\left\{\int_{\hat{\theta}-\varphi_n n^{-1/2}}^{\hat{\theta}+\varphi_n n^{-1/2}} \pi(\zeta)\exp\{lr_e(\zeta)\}d\zeta\right\}\right] \leq -E\left[\log\left\{\int_{\hat{\theta}-\varphi_n n^{-1/2}}^{\hat{\theta}+\varphi_n n^{-1/2}} \pi(\zeta)\exp\{lr(\zeta)\}I\{B_n(\zeta)\}d\zeta\right\}\right]$$

$$= -E\left[\log\left\{\int_{\hat{\theta}-\varphi_n n^{-1/2}}^{\hat{\theta}+\varphi_n n^{-1/2}} \pi(\zeta)\exp\{lr(\zeta)\}d\zeta - \int_{\hat{\theta}-\varphi_n n^{-1/2}}^{\hat{\theta}+\varphi_n n^{-1/2}} \pi(\zeta)\exp\{lr(\zeta)\}I\{B_n^c(\zeta)\}d\zeta\right\}\right].$$

Thus, defining $L_n = \int_{\hat{\theta}-\varphi_n n^{-1/2}}^{\hat{\theta}+\varphi_n n^{-1/2}} \pi(\zeta)\exp\{lr(\zeta)\}I\{B_n^c(\zeta)\}d\zeta$, one can rewrite the inequality above as

$$-E\left[\log\left\{\int_{\hat{\theta}-\varphi_n n^{-1/2}}^{\hat{\theta}+\varphi_n n^{-1/2}} \pi(\zeta)\exp\{lr_e(\zeta)\}d\zeta\right\}\right]$$

$$\leq -E\left[\log\left\{\int_{\hat{\theta}-\varphi_n n^{-1/2}}^{\hat{\theta}+\varphi_n n^{-1/2}} \pi(\zeta)\exp\{lr(\zeta)\}d\zeta - L_n\right\}I\{L_n \leq \exp(-n)\}\right]$$

$$-E\left[\log\left\{\int_{\hat{\theta}-\varphi_n n^{-1/2}}^{\hat{\theta}+\varphi_n n^{-1/2}} \pi(\zeta)\exp\{lr(\zeta)\}d\zeta - L_n\right\}I\{L_n > \exp(-n)\}\right]$$

$$\leq -E\left[\log\left\{\int_{\hat{\theta}-\varphi_n n^{-1/2}}^{\hat{\theta}+\varphi_n n^{-1/2}} \pi(\zeta)\exp\{lr(\zeta)\}d\zeta - \exp(-n)\right\}\right]$$

$$-E\left[\left\{\log\left(\int_{\hat{\theta}-\varphi_n n^{-1/2}}^{\hat{\theta}+\varphi_n n^{-1/2}} \pi(\zeta)\exp\{lr(\zeta)\}I\{B_n(\zeta)\}d\zeta\right) - J_n\right\}I\{L_n > \exp(-n)\}\right], \quad (S.10)$$

where $J_n = \log\left(\int_{\hat{\theta}-\varphi_n n^{-1/2}}^{\hat{\theta}+\varphi_n n^{-1/2}} \pi(\zeta)\exp\{lr(\zeta)\}d\zeta - \exp(-n)\right)$.

Next we show that the remainder term in (S.10) satisfies

$$-E\left[\left\{\log\left(\int_{\hat{\theta}-\varphi_n n^{-1/2}}^{\hat{\theta}+\varphi_n n^{-1/2}} \pi(\zeta)\exp\{lr(\zeta)\}I\{B_n(\zeta)\}d\zeta\right) - J_n\right\}I\{L_n > \exp(-n)\}\right] = o(1).$$

Towards this end, we apply the Cauchy-Schwarz inequality and Lemma A6 to obtain

$$\left|E\left[\left\{\log\left(\int_{\hat{\theta}-\varphi_n n^{-1/2}}^{\hat{\theta}+\varphi_n n^{-1/2}} \pi(\zeta)\exp\{lr(\zeta)\}I\{B_n(\zeta)\}d\zeta\right) - J_n\right\}I\{L_n > \exp(-n)\}\right]\right|$$

$$\leq \left[E\left\{\log\left(\int_{\hat{\theta}-\varphi_n n^{-1/2}}^{\hat{\theta}+\varphi_n n^{-1/2}} \pi(\zeta)\exp\{lr(\zeta)\}I\{B_n(\zeta)\}d\zeta\right) - J_n\right\}^2\right]^{1/2} [\Pr\{L_n > \exp(-n)\}]^{1/2}$$

$$= O(n^{4/3-2\beta})[\Pr\{L_n > \exp(-n)\}]^{1/2}, \quad (S.11)$$

where $\beta \in (0, 1/6)$ is defined in Lemma A6. According to (6), $B_n^c(\zeta)$ is defined as

$$I\{B_n^c(\zeta)\} = I\left\{\sum_{i=1}^n G(X_i,\zeta)^2 I(G(X_i,\zeta) < 0) < M \text{ or } \sum_{i=1}^n G(X_i,\zeta)^2 I(G(X_i,\zeta) > 0) < M\right\}.$$

This implies that $\Pr\{L_n > \exp(-n)\}$ in (S.11) satisfies

$$\Pr\{L_n > \exp(-n)\}$$

$$\leq \Pr\left\{\int_{\hat{\theta}-\varphi_n n^{-1/2}}^{\hat{\theta}+\varphi_n n^{-1/2}} \pi(\zeta)\exp\{lr(\zeta)\}I\left\{\sum_{i=1}^{n} G(X_i,\zeta)^2 I(G(X_i,\zeta)>0) < M\right\}d\zeta > \exp(-n)\right\}$$

$$+ \Pr\left\{\int_{\hat{\theta}-\varphi_n n^{-1/2}}^{\hat{\theta}+\varphi_n n^{-1/2}} \pi(\zeta)\exp\{lr(\zeta)\}I\left\{\sum_{i=1}^{n} G(X_i,\zeta)^2 I(G(X_i,\zeta)<0) < M\right\}d\zeta > \exp(-n)\right\}.$$

Without loss of generality, with respect to the assumptions of Proposition 1, we assume that $\partial G(X_i,\theta)/\partial\theta < 0$ for all $i=1,\ldots,n$. Then, for $\zeta \in [\hat{\theta}-\varphi_n n^{-1/2}, \hat{\theta}+\varphi_n n^{-1/2}]$, we obtain

$$\sum_{i=1}^{n} G(X_i,\zeta)^2 I\{G(X_i,\zeta)>0\} > \sum_{i=1}^{n} G(X_i,\zeta)^2 I\{G(X_i,\hat{\theta}+\varphi_n n^{-1/2})>0\} \quad \text{(S.12)}$$

and the first derivative of $\sum_{i=1}^{n} G(X_i,\zeta)^2 I\{G(X_i,\hat{\theta}+\varphi_n n^{-1/2})>0\}$ satisfies

$$2\sum_{i=1}^{n} G(X_i,\zeta)\frac{\partial G(X_i,\zeta)}{\partial \zeta} I\{G(X_i,\hat{\theta}+\varphi_n n^{-1/2})>0\} < 0. \quad \text{(S.13)}$$

The result (S.13) implies that $\sum_{i=1}^{n} G(X_i,\zeta)^2 I\{G(X_i,\hat{\theta}+\varphi_n n^{-1/2})>0\}$ decreases for $\zeta \in [\hat{\theta}-\varphi_n n^{-1/2}, \hat{\theta}+\varphi_n n^{-1/2}]$. Then by (S.12), we conclude with

$$\sum_{i=1}^{n} G(X_i,\zeta)^2 I(G(X_i,\zeta)>0) > \sum_{i=1}^{n} G(X_i,\hat{\theta}+\varphi_n n^{-1/2})^2 I\{G(X_i,\hat{\theta}+\varphi_n n^{-1/2})>0\}.$$

Defining $\theta_1 = \hat{\theta}+\varphi_n n^{-1/2}$ and using that $lr(\zeta)<0$, $\int_{-\infty}^{\infty} \pi(\zeta)d\zeta = 1$, we have

$$\Pr\left[\int_{\hat{\theta}-\varphi_n n^{-1/2}}^{\hat{\theta}+\varphi_n n^{-1/2}} \pi(\zeta)\exp\{lr(\zeta)\}I\left\{\sum_{i=1}^{n} G(X_i,\zeta)^2 I(G(X_i,\zeta)>0) < M\right\}d\zeta > \exp(-n)\right]$$

$$\leq \Pr\left[\int_{\hat{\theta}-\varphi_n n^{-1/2}}^{\hat{\theta}+\varphi_n n^{-1/2}} \pi(\zeta)\exp\{lr(\zeta)\}I\left\{\sum_{i=1}^{n} G(X_i,\theta_1)^2 I(G(X_i,\theta_1)>0) < M\right\}d\zeta > \exp(-n)\right]$$

$$\leq \Pr\left[I\left\{\sum_{i=1}^{n} G(X_i,\theta_1)^2 I(G(X_i,\theta_1)>0) < M\right\}\int_{\hat{\theta}-\varphi_n n^{-1/2}}^{\hat{\theta}+\varphi_n n^{-1/2}} \pi(\zeta)\exp\{lr(\zeta)\}d\zeta > \exp(-n)\right]$$

$$\leq \Pr\left\{\sum_{i=1}^{n} G(X_i,\theta_1)^2 I(G(X_i,\theta_1)>0) < M\right\}.$$

In a similar manner to the analysis above, defining $\theta_2 = \hat{\theta}-\varphi_n n^{-1/2}$ one can show that

$$\Pr\left[\int_{\hat{\theta}-\varphi_n n^{-1/2}}^{\hat{\theta}+\varphi_n n^{-1/2}} \pi(\zeta)\exp\{lr(\zeta)\}I\left\{\sum_{i=1}^{n} G(X_i,\zeta)^2 I(G(X_i,\zeta)<0) < M\right\}d\zeta > \exp(-n)\right]$$

$$\leq \Pr\left\{\sum_{i=1}^{n} G(X_i,\theta_2)^2 I(G(X_i,\theta_2)<0) < M\right\}.$$

Taking into account the Taylor expansion $G(X_i,\theta_1) = G(X_i,\theta_0) + (\theta_1 - \theta_0)\partial G(X_i,\breve{\theta}_i)/\partial\theta$, where $\theta_0 = \theta + \varphi_n n^{-1/2}$, $\theta$ satisfies $E\{G(X_1,\theta)\} = 0$, $\breve{\theta}_i = \theta_0 + \breve{w}_i(\theta_1 - \theta_0)$ with $\breve{w}_i \in (0,1)$, and $|\partial G(X_i,\breve{\theta}_i)/\partial\theta| < Q(X_i)$ with $E\{Q(X_i)^4\} < \infty$, for all $i=1,\ldots,n$, one can then use Chebyshev's inequality to arrive at the expression $\Pr\left\{\sum_{i=1}^{n} G(X_i,\theta_1)^2 I(G(X_i,\theta_1)>0) < M\right\} = O(n^{-r/2})$, where

$$E\left|\sum_{i=1}^{n}\{G(X_i,\theta_1)^2 I(G(X_i,\theta_1)>0) - q\}\right|^r < \infty \quad \text{with} \quad q = E\{G(X_i,\theta_1)^2 I(G(X_i,\theta_1)>0)\} \quad \text{and}$$

$r > 16/3 - 8\beta$, since $E|G(X_1,\theta)|^{8+\gamma} < \infty$ as required in Lemma A8 statement. In a similar manner, one can show that $\Pr\left\{\sum_{i=1}^{n} G(X_i,\theta_2)^2 I(G(X_i,\theta_2)<0) < M\right\} = O(n^{-r/2})$, where

$$q_2 = E\{G(X_i,\theta_2)^2 I(G(X_i,\theta_2)<0)\} \quad \text{and} \quad E\left|\sum_{i=1}^{n}\{G(X_i,\theta_2)^2 I(G(X_i,\theta_2)<0) - q_2\}\right|^r < \infty.$$

Thus by virtue of (S.11) and the above analysis, we conclude with

$$-E\left[\left\{\log\left(\int_{\hat{\theta}-\varphi_n n^{-1/2}}^{\hat{\theta}+\varphi_n n^{-1/2}} \pi(\zeta)\exp\{lr(\zeta)\}I\{B_n(\zeta)\}d\zeta\right) - J_n\right\}I\{L_n > \exp(-n)\}\right] = o(1). \quad (S.14)$$

By (S.10) and (S.14), we obtain

$$-E\left[\log\left\{\int_{\hat{\theta}-\varphi_n n^{-1/2}}^{\hat{\theta}+\varphi_n n^{-1/2}} \pi(\zeta)\exp\{lr_e(\zeta)\}d\zeta\right\}\right]$$

$$\leq -E\left[\log\left\{\int_{\hat{\theta}-\varphi_n n^{-1/2}}^{\hat{\theta}+\varphi_n n^{-1/2}} \pi(\zeta)\exp\{lr(\zeta)\}d\zeta - \exp(-n)\right\}\right] + o(1). \quad (S.15)$$

Applying a Taylor expansion to the function $\log\left\{\int_{\hat{\theta}-\varphi_n n^{-1/2}}^{\hat{\theta}+\varphi_n n^{-1/2}} \pi(\zeta)\exp\{lr(\zeta)\}d\zeta - s\right\}$ of $s = \exp(-n)$ about 0, we can rewrite (S.15) as

$$-E\left[\log\left\{\int_{\hat{\theta}-\varphi_n n^{-1/2}}^{\hat{\theta}+\varphi_n n^{-1/2}} \pi(\zeta)\exp\{lr_e(\zeta)\}d\zeta\right\}\right]$$

$$\leq -E\left[\log\left\{\int_{\hat{\theta}-\varphi_n n^{-1/2}}^{\hat{\theta}+\varphi_n n^{-1/2}}\pi(\zeta)\exp\{lr(\zeta)\}d\zeta\right\}\right] + E\left[\exp(-n)\left\{\int_{\hat{\theta}-\varphi_n n^{-1/2}}^{\hat{\theta}+\varphi_n n^{-1/2}}\pi(\zeta)\exp\{lr(\zeta)\}d\zeta - e_R\right\}^{-1}\right] + o(1),$$

(S.16)

where $e_R \in (0, \exp(-n))$.

To complete the proof of Lemma A8, we show the following two results regarding components of (S.16)

i) $-E\left[\log\left\{\int_{\hat{\theta}-\varphi_n n^{-1/2}}^{\hat{\theta}+\varphi_n n^{-1/2}}\pi(\zeta)\exp\{lr(\zeta)\}d\zeta\right\}\right] \leq -\log\left[\pi(\theta)\{2\pi\sigma^2(\theta)/n\}^{1/2}\right] + o(1);$

ii) $E\left[\exp(-n)\left\{\int_{\hat{\theta}-\varphi_n n^{-1/2}}^{\hat{\theta}+\varphi_n n^{-1/2}}\pi(\zeta)\exp\{lr(\zeta)\}d\zeta - e_R\right\}^{-1}\right] = o(1).$

The evaluation of (*i*) is based on the following scheme. Considering the Taylor expansion of $lr(\theta)$, we have

$$lr(\theta) = lr(\hat{\theta}) + (\theta - \hat{\theta})\frac{\partial lr(\hat{\theta})}{\partial \theta} + \frac{1}{2}(\theta - \hat{\theta})^2 \frac{\partial^2 lr(\hat{\theta})}{\partial \theta^2} + \frac{1}{6}(\theta - \hat{\theta})^3\left(\frac{\partial^3 lr(u)}{\partial \theta^3}\bigg|_{u=\theta+\varpi(\hat{\theta}-\theta)}\right), \quad \varpi \in (0,1). \quad (S.17)$$

Since $lr(\theta) = -\sum_{i=1}^n \log\{n + \lambda G(X_i, \theta)\} - \log(n^{-n})$, one can show that

$$\frac{\partial lr(\theta)}{\partial \theta} = -\lambda(\theta)\sum_{i=1}^n \frac{\partial G(X_i,\theta)/\partial \theta}{n + \lambda(\theta)G(X_i,\theta)},$$

$$\frac{\partial^2 lr(\theta)}{\partial \theta^2} = -\frac{\partial \lambda(\theta)}{\partial \theta}\sum_{i=1}^n \frac{\partial G(X_i,\theta)/\partial \theta}{n + \lambda(\theta)G(X_i,\theta)} - \lambda(\theta)A(\theta),$$

$$\frac{\partial^3 lr(\theta)}{\partial \theta^3} = -\frac{\partial^2 \lambda(\theta)}{\partial \theta^2}\sum_{i=1}^n \frac{\partial G(X_i,\theta)/\partial \theta}{n + \lambda G(X_i,\theta)} - 2\frac{\partial \lambda(\theta)}{\partial \theta}A(\theta) - \lambda(\theta)\frac{\partial A(\theta)}{\partial \theta}, \quad (S.18)$$

$$\frac{\partial \lambda(\theta)}{\partial \theta} = \frac{n\sum_{i=1}^n p_i^2 \partial G(X_i,\theta)/\partial \theta}{\sum_{i=1}^n p_i^2 G(X_i,\theta)^2}, \quad (S.19)$$

$$\frac{\partial^2 \lambda(\theta)}{\partial \theta^2} = \left[\sum_{i=1}^n p_i^2 G(X_i,\theta)^2\right]^{-1}\left[\sum_{i=1}^n p_i^2 \left\{n(\partial^2 G(X_i,\theta)/\partial \theta^2) - 2(\partial \lambda(\theta)/\partial \theta)(\partial G(X_i,\theta)/\partial \theta)G(X_i,\theta)\right\}\right.$$

$$-2\sum_{i=1}^{n} p_i^3 \{n(\partial G(X_i,\theta)/\partial \theta)-(\partial \lambda(\theta)/\partial \theta)G(X_i,\theta)^2\}\{(\partial \lambda(\theta)/\partial \theta)G(X_i,\theta)+\lambda(\theta)(\partial G(X_i,\theta)/\partial \theta)\}\Big]$$

(S.20)

where $\lambda(\theta)$ satisfies $\sum_{i=1}^{n} G(X_i,\theta)/\{n+\lambda(\theta)G(X_i,\theta)\}=0$, $p_i = \{n+\lambda G(X_i,\theta)\}^{-1}$, $i=1,...,n$, and

$$A(\theta) = \sum_{i=1}^{n} \frac{\partial^2 G(X_i,\theta)/\partial \theta^2}{\{n+\lambda G(X_i,\theta)\}^2} - \frac{\partial \lambda(\theta)}{\partial \theta}\sum_{i=1}^{n} \frac{G(X_i,\theta)\partial G(X_i,\theta)/\partial \theta}{\{n+\lambda G(X_i,\theta)\}^2} - \lambda(\theta)\sum_{i=1}^{n} \frac{(\partial G(X_i,\theta)/\partial \theta)^2}{\{n+\lambda G(X_i,\theta)\}^2},$$

for details see Vexler, Hutson, and Yu (2014b).

Thus

$$lr(\hat{\theta}) = \partial lr(\hat{\theta})/\partial \theta = \lambda(\hat{\theta}) = 0,$$

$$\frac{\partial \lambda(\theta)}{\partial \theta}\bigg|_{\theta=\hat{\theta}} = \frac{\sum_{i=1}^{n} \partial G(X_i,\theta)/\partial \theta|_{\theta=\hat{\theta}}}{n^{-1}\sum_{i=1}^{n} G(X_i,\hat{\theta})^2}, \quad \frac{\partial^2 lr(\hat{\theta})}{\partial \theta^2} = -\frac{n^{-1}\left\{\sum_{i=1}^{n} \partial G(X_i,\theta)/\partial \theta|_{\theta=\hat{\theta}}\right\}^2}{n^{-1}\sum_{i=1}^{n} G(X_i,\hat{\theta})^2}, \quad \text{(S.21)}$$

since $\hat{\theta}$ maximizes $lr(\theta)$ as well as $n^{-1}\sum_{i=1}^{n} G(X_i,\hat{\theta}) = 0$.

It turns out that there are the following rules: 1) $\partial^3 lr(\theta)/\partial \theta^3$ depends on $\partial^2 \lambda(\theta)/\partial \theta^2$; 2) $\partial \lambda(\theta)/\partial \theta$ and $\lambda(\theta)$ via (S.18); 3) $\partial^2 \lambda(\theta)/\partial \theta^2$ depends on $\partial \lambda(\theta)/\partial \theta$ and $\lambda(\theta)$ via (S.20); and 4) $\partial \lambda(\theta)/\partial \theta$ depends on $\lambda(\theta)$ via (S.19), where $\lambda(\theta) = O_p(n^{2/3})$ by Lemma A7. This leads to the conclusion that $\partial^3 lr(\theta)/\partial \theta^3 = O_p(n)$ for $\theta \in [\hat{\theta}-\varphi_n n^{-1/2}, \hat{\theta}+\varphi_n n^{-1/2}]$. This result is confirmed by Lemma 7 of Zhong and Ghosh (2016, p. 3033) that provides $|\partial^3 lr(\theta)/\partial \theta^3| \leq C_1 n$ almost surely for relatively large $n$, where $C_1$ denotes a positive constant. Defining $\Lambda = \max_{\theta \in [\hat{\theta}-\varphi_n n^{-1/2}, \hat{\theta}+\varphi_n n^{-1/2}]} |\partial^3 lr(\theta)/\partial \theta^3|$, it is clear that the result above and (S.17) imply the following inequality regarding the component $-E\left[\log\left\{\int_{\hat{\theta}-\varphi_n n^{-1/2}}^{\hat{\theta}+\varphi_n n^{-1/2}} \pi(\zeta)\exp\{lr(\zeta)\}d\zeta\right\}\right]$ at (S.16),

$$-E\left[\log\left\{\int_{\hat{\theta}-\varphi_n n^{-1/2}}^{\hat{\theta}+\varphi_n n^{-1/2}} \pi(\zeta)\exp\{lr(\zeta)\}d\zeta\right\}\right]$$

$$\leq -E\left[\log\left\{\int_{\hat{\theta}-\varphi_n n^{-1/2}}^{\hat{\theta}+\varphi_n n^{-1/2}} \pi(\zeta)\exp\left\{-\frac{1}{2}\frac{\partial^2 lr(\hat{\theta})}{\partial \zeta^2}(\zeta-\hat{\theta})^2 - \Lambda \varphi_n^3 n^{-1.5}\right\}d\zeta\right\}\right]$$

$$= -E\left[\log\left\{\int_{\hat{\theta}-\varphi_n n^{-1/2}}^{\hat{\theta}+\varphi_n n^{-1/2}} \pi(\zeta)\exp\left\{-\frac{1}{2}\frac{\partial^2 lr(\hat{\theta})}{\partial \zeta^2}(\zeta-\hat{\theta})^2\right\}d\zeta\right\}\right] - \varphi_n^3 n^{-1.5} E(\Lambda)$$

$$= -E\left[\log\left\{\pi(\hat{\theta})\int_{-\infty}^{+\infty}\exp\left\{-\frac{1}{2}\frac{\partial^2 lr(\hat{\theta})}{\partial \zeta^2}(\zeta-\hat{\theta})^2\right\}d\zeta\right\}\right] + o(1)$$

$$= -E\left[\log\left\{\pi(\hat{\theta})(2\pi\sigma_M^2/n)^{1/2}\right\}\right] + o(1) = -\log\left[\pi(\theta)\{2\pi\sigma^2(\theta)/n\}^{1/2}\right] + o(1), \qquad (S.22)$$

where $\sigma^2(\theta) = E\{G(X,\theta)\}^2 / \{EG'(X,\theta)\}^2$ and $G'(X,\theta) = \partial G(X,\theta)/\partial\theta$.

The above analysis shows the statement (*i*) regarding (S.16) is in effect. Next we proceed to prove statement (*ii*) regarding (S.16).

By virtue of (S.17) and the analysis above, it is obvious that $\int_{\hat{\theta}-\varphi_n n^{-1/2}}^{\hat{\theta}+\varphi_n n^{-1/2}} \pi(\zeta)\exp\{lr(\zeta)\}d\zeta > \exp(-n)$ almost surely, since $\int_{\hat{\theta}-\varphi_n n^{-1/2}}^{\hat{\theta}+\varphi_n n^{-1/2}} \pi(\zeta)\exp\{lr(\zeta)\}d\zeta = O(n^{-1/2})$ almost surely for relatively large *n*. Taking into account that $lr(\hat{\theta}) = 0$, and then $\int_{\hat{\theta}-\varphi_n n^{-1/2}}^{\hat{\theta}+\varphi_n n^{-1/2}} \pi(\zeta)\exp\{lr(\zeta)\}d\zeta \geq \pi(\hat{\theta})$, we have, for $e_R \in (0, \exp(-n))$,

$$\frac{\exp(-n)}{\int_{\hat{\theta}-\varphi_n n^{-1/2}}^{\hat{\theta}+\varphi_n n^{-1/2}} \pi(\zeta)\exp\{lr(\zeta)\}d\zeta - e_R} \cong \frac{\exp(-n)}{\int_{\hat{\theta}-\varphi_n n^{-1/2}}^{\hat{\theta}+\varphi_n n^{-1/2}} \pi(\zeta)\exp\{lr(\zeta)\}d\zeta} \leq \exp(-n)\{\pi(\hat{\theta})\}^{-1}. \quad (S.23)$$

Then (S.23) implies (*ii*) regarding (S.16) in the form of

$$E\left[\exp(-n)\left\{\int_{\hat{\theta}-\varphi_n n^{-1/2}}^{\hat{\theta}+\varphi_n n^{-1/2}} \pi(\zeta)\exp\{lr(\zeta)\}d\zeta - e_R\right\}^{-1}\right] = o(1).$$

Thus taking into account the above result and (S.22), we obtain that

$$-E\left[\log\left\{\int_{\hat{\theta}-\varphi_n n^{-1/2}}^{\hat{\theta}+\varphi_n n^{-1/2}} \exp\{lr_e(\zeta)\}\pi(\zeta)d\zeta\right\}\right] \leq -\log\left[\pi(\theta)\{2\pi\sigma^2(\theta)/n\}^{1/2}\right] + o(1).$$

In a similar manner to the proof scheme above with respect to (S.9), one can show

$$-E\left[\log\left\{\int_{\hat{\theta}-\varphi_n n^{-1/2}}^{\hat{\theta}+\varphi_n n^{-1/2}}\exp\{lr_e(\zeta)\}\pi(\zeta)d\zeta+d_n\right\}\right]\geq-\log\left[\pi(\theta)\{2\pi\sigma^2(\theta)/n\}^{1/2}\right]+o(1).$$

This completes the proof of Lemma A8.

## References

Clarke, B. and Yuan, A. 2010. Reference priors for empirical likelihoods, In Chen, M. H., Mueller, P., Sun, D., Ye, K. & Dey, D. K. (eds.), *Frontiers of Statistical Decision Making and Bayesian Analysis: In honor of James O. Berger*. 56–68. New York: Springer.

DiCiccio, T., Hall, P., and Romano, J. 1991. Empirical likelihood is Bartlett-correctable, *The Annals of Statistics,* 19, 1053-1061.

Hartigan, J. A. 1998. The maximum likelihood prior, *The Annals of Statistics,* 26, 2083-2103.

Owen, A. 1990. Empirical likelihood ratio confidence regions, *The Annals of Statistics,* 18, 90-120.

Owen, A. 2001. *Empirical Likelihood*, New York: Chapman and Hall.

Vexler, A., Liu, S. L., Kang, L., and Hutson, A. D. 2009. Modifications of the Empirical Likelihood Interval Estimation with Improved Coverage Probabilities, Communications in Statistics, Simulation and Computation, 38, 2171-2183.

Vexler, A., Tao, G., and Hutson, A. D. 2014a. Posterior expectation based on empirical likelihoods, *Biometrika,* 101, 711-718.

Vexler, A, Hutson, A. D. and Yu, J. 2014b. Empirical likelihood methods in clinical experiments, (In Balakrishman, N. eds.), *Methods and Applications of Statistics in Clinical Trials: Encyclopedia of Clinical Trials*. John Wiley & Sons, Newark, NJ

Zhong, X. L., and Ghosh, M. 2016. Higher-order properties of Bayesian empirical likelihood, *Electronic Journal of Statistics,* 10, 3011-3044.